\documentclass{pasj01}
\draft 
\Received{2022 June 23}
\Accepted{2022 November 7}
\Published{$\langle$publication date$\rangle$}

\usepackage{lineno}

\begin{document}

\title{
Modeling of geocoronal solar wind charge exchange events detected with Suzaku 
}\author{
Daiki \textsc{Ishi},\altaffilmark{1,}$^{*}$
Kumi \textsc{Ishikawa},\altaffilmark{1}
Yoshizumi \textsc{Miyoshi},\altaffilmark{2}
Naoki \textsc{Terada},\altaffilmark{3} and 
Yuichiro \textsc{Ezoe}\altaffilmark{1}
}
\altaffiltext{1}{Department of Physics, Tokyo Metropolitan University, 1-1 Minami-Osawa, Hachioji, Tokyo 192-0397, Japan}
\altaffiltext{2}{Institute for Space-Earth Environmental Research, Nagoya University, Furo-cho, Chikusa-ku, Nagoya, Aichi 464-8601, Japan}
\altaffiltext{3}{Department of Geophysics, Tohoku University, 6-3 Aramaki-Aza-Aoba, Aoba-ku, Sendai, Miyagi 980-8578, Japan}
\email{ishi-daiki@ed.tmu.ac.jp}


\maketitle

\begin{abstract}
A model of geocoronal solar wind charge exchange (SWCX) emission was built and 
compared to five Suzaku detections of bright geocoronal SWCX events. 
An exospheric neutral hydrogen distribution model, 
charge exchange cross sections, 
solar wind ion data taken with the ACE and WIND satellites, 
and magnetic field models of the Earth's magnetosphere 
are all combined in order to predict time-variable geocoronal SWCX emission 
depending on line-of-sight directions of the Suzaku satellite. 
The modeled average intensities of O\,\emissiontype{VII} emission lines were consistent 
with the observed ones within a factor of three in four out of the five cases 
except for an event in which a line-of-sight direction was toward the night side of the high-latitude magnetosheath and 
a major geomagnetic storm was observed. 
Those of O\,\emissiontype{VIII} emission lines were underestimated by a factor of three or more in all the five cases. 
On the other hand, the modeled O\,\emissiontype{VII} and O\,\emissiontype{VIII} light curves reproduced the observed ones
after being scaled by ratios between the observed and modeled average intensities. 
In particular, short-term variations due to line-of-sight directions traversing cusp regions 
during an orbital motion of the Suzaku satellite were reproduced. 
These results are discussed in the context of model uncertainties. 
\end{abstract}


\section{Introduction}

Solar wind charge exchange (SWCX) is a ubiquitous phenomenon throughout our solar system. 
Highly charged solar wind ions 
acquire an electron from neutral atoms or molecules. 
The electron first enters into an excited state but soon cascades to the ground state through all possible transitions 
by producing emission lines in the extreme ultraviolet to soft X-ray energy range. 
This process was first proposed to explain a copious flux of soft X-rays 
from comet Hyakutake \citep{lis96,cra97}. 
Similar emissions have been observed from a number of other comets 
and solar system planets (see \cite{bha07} for review). 


The same process operates to produce soft X-rays 
in the Earth's exosphere or geocorona and in the heliosphere. 
These two phenomena are called geocoronal and heliospheric SWCX. 
The former arises from solar wind plasma mostly in the Earth's magnetosheath 
interacting with exospheric neutrals extending more than 10 Earth radii ($R_\mathrm{E}$), 
while the latter from that interacting with interstellar neutrals 
permeating the heliosphere whose radius is about 100 AU. 
These emissions contaminate signals of interest as temporally variable backgrounds, 
e.g., enhanced backgrounds with time scales of several hours to a couple of days 
as observed during the ROSAT All-Sky Survey (e.g., \cite{sno94}). 
Such an astrophysical nuisance has also been recognized as a severe problem for other X-ray astronomy satellites 
sensitive to soft X-rays (e.g., Chandra, XMM-Newton, and Suzaku). 


All the soft X-ray observations from Earth-orbiting satellites must contend with 
a persistent foreground emission from both geocoronal and heliospheric SWCX
(see \cite{kun19} for more complete discussions of problems posed to astrophysics). 
The former is typically an order of magnitude weaker than the latter 
but responds much more quickly (on time scales of less than an hour) to abrupt solar wind changes, 
thereby producing a sporadic contamination 
that sometimes reaches the same order of magnitude as the latter or even greater. 
The latter varies on much longer time scales and more subtle variations. 
To determine each contribution to astronomical observations, careful checks of background signals 
combined with simultaneous solar wind observations are indispensable. 


Recent X-ray astronomical observations with X-ray CCDs onboard Chandra, XMM-Newton, and Suzaku 
have detected numerous SWCX events (e.g., \cite{war04,sno04,fuj07}). 
\citet{car08} and \citet{car11} systematically searched for geocoronal SWCX events from XMM-Newton archival data. 
Most of the events occurred during periods when XMM-Newton observed through the sub-solar side of the magnetosheath. 
Models to simulate a spatial distribution of geocoronal SWCX emission (e.g., \cite{rob06}) indicate that 
the strongest emitters are present in the nose of the magnetosheath and in the magnetospheric cusps 
owing to dense populations of solar wind plasma and exospheric neutrals. 
Strong SWCX events were sometimes detected even when their line-of-sight directions intersected 
regions where no strong emissions are expected (e.g., the flanks of the magnetosheath). 
These cases might originate from heliospheric SWCX events such as 
coronal mass ejections (CMEs) passing through local heliospheric structures (e.g., \cite{kou07}). 
\citet{car10} argued that, although XMM-Newton was not pointed to the regions where the strongest emissions are expected, 
i.e., neither the nose of the magnetosheath nor the magnetospheric cusps, 
the strongest, the most spectrally rich case was attributed to a CME passing through the Earth on 2001 October 21. 


\citet{car11} attempted to model roughly a hundred XMM-Newton observations 
contaminated with time-variable geocoronal SWCX emission 
using a set of models to compute positions of the magnetosheath boundaries, 
i.e., the bow shock and the magnetopause, and 
to predict solar wind conditions in the near-Earth region (e.g., \cite{spr66}). 
Approximately 50\% of the modeled fluxes agreed with the observed ones within a factor of two. 
\citet{whi16} compared results from an magnetohydrodynamics (MHD)-based model 
with 19 strong SWCX events listed in \citet{car11}, 
giving only 6 cases (approximately 30\%) returning count rates within a factor of two of the observed values. 


The bulk of these studies attempting to model geocoronal SWCX emission observed with Chandra and XMM-Newton 
have been concerned with observations of the bright nose of the magnetosheath from 
positions that maximize the path length through the magnetosheath (see \cite{kun15}). 
Models of the magnetosheath have significant difficulties determining distances 
from the Earth to the noses of the bow shock and the magnetopause. 
The observer's line of sight might have completely missed the bright nose of the magnetosheath 
even with an error of 1 $R_\mathrm{E}$, which is one of the main reasons for large discrepancies between 
models and observations. 
This problem has also been exemplified by comparison of an MHD model 
with a dozen Chandra SWCX observations \citep{war14}. 


The X-ray Imaging Spectrometer (XIS; \cite{koy07}) onboard Suzaku \citep{mit07} is 
one of the best instruments to observe geocoronal SWCX emission 
thanks to its low and stable background rate and good energy resolution and response. 
\citet{fuj07} and \citet{ezo10} detected enhanced background events 
in the directions of the north ecliptic pole and the celestial equator, respectively. 
These enhancements showed significant temporal correlations with 
simultaneously observed solar wind fluxes, 
which is one of the most explicit signatures of geocoronal SWCX events. 
\citet{ezo11} and \citet{ish13} reported strong SWCX events associated with 
arrivals of increased solar wind fluxes during intense geomagnetic storms. 
\citet{ish19} found an event in which background signals tracked abrupt solar wind changes 
due to a CME-induced interplanetary shock and the CME itself. 
These observed emissions exceeded by an order of magnitude compared to those predicted from 
upstream solar wind fluxes and exospheric neutral hydrogen column densities (e.g., \cite{ezo10,ezo11,ish13}). 


Compared to observations with Chandra and XMM-Newton, 
Suzaku can observe geocoronal SWCX emission through line-of-sight directions roughly perpendicular to 
the surface of the magnetosheath, i.e., the flanks of the magnetosheath. 
Model uncertainties in the nose of the magnetosheath have smaller effects on their results, 
e.g., overall fluxes and light curves. 
Modeling difficulties stem from the fact that the path length through the magnetosheath is much shorter and 
geocoronal SWCX emission from the flanks of the magnetosheath is intrinsically weaker. 


The low-Earth orbit of Suzaku allows us to observe one of the strongest emitters, i.e., the magnetospheric cusps, 
which is difficult to observe with the high-Earth orbit satellites (e.g., Chandra and XMM-Newton) 
due to Earth avoidance angles. 
Temporal variations with time scales of several minutes as observed with Suzaku (e.g., \cite{fuj07}) are attributed to 
abrupt changes in the path length through the near-cusp region during an orbital motion of Suzaku. 
There are no good models for such geocoronal SWCX events. 


In this paper, making the most of the instrumental and orbital advantages of Suzaku, 
we attempt to model bright geocoronal SWCX events detected with Suzaku. 
To accurately compare model results, we re-analyze these events in the same manner as \citet{ezo11}. 
Unlike our previous predictions using time-averaged solar wind fluxes 
and roughly assumed bow shock and magnetopause positions 
depending on line-of-sight directions (e.g., \cite{ezo10,ezo11,ish13}), 
our new model incorporates time-variable solar wind element and ion abundances 
measured by upstream solar wind monitoring satellites 
and empirical bow shock and magnetopause models. 
Our model also takes into account cusp geometries by tracing magnetic field lines, thereby capable of predicting 
light curves of geocoronal SWCX emission through the near-cusp region associated with the low-Earth orbit. 
Below we first re-analyze the five bright SWCX events, 
describe the construction of our model, and then 
compare model results with observational data. 

\section{Observational Data}

Our observational data consist of the five Suzaku detections of bright geocoronal SWCX events listed in Table \ref{tab1}. 
These observations showed clear temporal variations in soft X-ray backgrounds, 
which are established as time-variable components of geocoronal SWCX emission 
in the past studies. 
Those of heliospheric SWCX emission are expected to be constant during observations of several hours to a couple of days. 
Therefore, we can extract only geocoronal SWCX emission. 


The observational parameters are summarized in Table \ref{tab1}. 
The solar activity during the first four observations (ObsIDs: 10009010, 100014010, 100018010, and 50009010) 
was approaching minimum, the end of the 23rd solar cycle, 
while that during the last observation (ObsID: 508072010) 
was around the maximum of the 24th solar cycle. 
Hereafter, we call these observations ID1--5 as defined in Table \ref{tab1}. 


There is another bright SWCX event detected with Suzaku as reported in \citet{asa21}. 
It consists of a set of two observations on 2005 September 11--13 and 2006 January 26--27. 
The former spectrum contained a series of enhanced emission lines from highly 
charged ions in CMEs, while the latter one did not. 
Their solar activities and line-of-sight directions relative to a stream of interstellar neutrals, i.e., 
the downwind gravitational focusing cone of interstellar helium atoms, are different, 
resulting in a potentially different heliospheric SWCX contribution to each observation. 
Considering significant difficulties predicting solar wind propagations and interstellar neutral distributions 
on large spatial scales, we did not use this event for comparison. 


Figure 1 shows average line-of-sight directions in Geocentric Solar Ecliptic (GSE) coordinates during each observation. 
These line-of-sight directions passed through a variety of regions in the Earth's magnetosphere, i.e., 
the dusk side of the mid-latitude magnetosheath (ID1), 
the night side of the high-latitude magnetosheath (ID2), 
the northern polar cusp (ID3), 
the day side of the low-latitude magnetosheath (ID4), 
and the southern polar cusp (ID5). 
The Suzaku solar-angle constraint had been tightened from 65$^\circ$--115$^\circ$ to 70$^\circ$--110$^\circ$ 
during the Suzaku AO-7 cycle (2012 April to 2013 March)
due to the solar panel degradation in power output.\footnote{
$\langle$https://heasarc.gsfc.nasa.gov/docs/suzaku/news/power.html$\rangle$.}
The former applied to the observations of ID1--4, while the latter to that of ID5. 
This change may influence scattering of solar X-rays to each observation. 
Therefore, we quantitatively checked this possibility later in subsection 3.2. 


The Suzaku/XIS instrument consists of three front-illuminated CCDs (XIS 0, 2, and 3) and 
one back-illuminated CCD (XIS 1). 
We used only XIS 1 data because it is more sensitive to soft X-rays than the other detectors. 
Figure \ref{fig2} shows an XIS 1 image in the 0.2--1 keV band during each observation. 
To minimize contamination from bright X-ray source(s) in the field of view of ID1, 2, and 5, 
we chose triangle or polygon regions located at the corners of each field of view 
for the following light curve and spectral analyses. 
Hereafter, we call these regions terrestrial diffuse X-ray (TDX) regions. 
For those of ID3 and 4, we defined a circular region with a radius of \timeform{8'.5} as the TDX region 
by considering well-calibrated radial profiles of contamination distributions on optical blocking filters. 
There are two sources emitting hard X-rays above 2 keV in the field of view of ID4 (see \cite{ebi08}). 
Therefore, we excluded these sources or two circular regions with radii of \timeform{2'} and \timeform{2'.5} 
from the TDX region. 
The total areas of each TDX region are summarized in Table \ref{tab1}. 

\section{Analysis}

\subsection{Data Reduction}

Using the HEAsoft version 6.27.2 package, we performed data analysis from cleaned event files 
screened through the Suzaku final pipeline processing version 3.0.22.43 and 3.0.22.44, 
both reprocessed after the end of the satellite operations in 2015.\footnote{
$\langle$https://heasarc.gsfc.nasa.gov/docs/suzaku/archive/suza\_procversion.html$\rangle$.}
Events were filtered by standard screening criteria,\footnote{
$\langle$https://heasarc.gsfc.nasa.gov/docs/suzaku/processing/criteria\_xis.html$\rangle$.}
which remove high-background intervals mainly during passages 
through the South Atlantic Anomaly and through regions of low geomagnetic cut-off-rigidities. 


Hot and flickering pixels were removed with the latest calibration database 
but the number of noise pixels cumulatively increased in the later phase of the Suzaku mission, 
resulting in an increased non-X-ray background (NXB) level. 
For the data of ID5 observed in 2013, 
we excluded noise events by a cumulative flickering pixel map 
identified by the calibration team of Suzaku/XIS.\footnote{
$\langle$https://heasarc.gsfc.nasa.gov/docs/suzaku/analysis/xisnxbnew.html$\rangle$.}
For those of ID1--4 observed in 2005, soon after the launch of the satellite, 
we did not exclude such events because such pixels were not identified at that time. 


Although XIS 1 kept a spectral resolution good enough in the early observation, 
it had become worse due to radiation damage by cosmic particles. 
To rejuvenate its spectral resolution by filling charge traps with artificially injected charges, 
spaced-row charge injection had been performed since 2006. 
For the observation of ID5, we removed not only the first but also second rows adjacent to charge injected ones 
to avoid leaked events due to an increased charge injection from 2 keV to 6 keV after 2011.\footnote{
$\langle$https://heasarc.gsfc.nasa.gov/docs/suzaku/analysis/nxb\_ci6kev.html$\rangle$.}


The above additional screening of ID5 resulted in an improved signal-to-noise ratio in soft bands below 1 keV. 
The effective area of the TDX region decreased by $\sim$6\%, while the NXB rate in the 0.2--1 keV band 
derived from the night-Earth database \citep{taw08} was suppressed by $\sim$19\%. 

\subsection{Removal of Scattered Solar X-rays} 

The interaction between solar X-rays and neutral oxygen atoms or molecules in the Earth's atmosphere produces 
a fluorescent emission line at 0.525 keV. 
This line sometimes appears even after excluding periods where 
elevation angles from the Earth rim (ELV) and the bright-Earth rim (DYE\_ELV) are 
less than 5$^\circ$ and 20$^\circ$, respectively (e.g., \cite{ezo11,sek14}). 


For each observation, we checked spectra extracted from the TDX region at different ELV values of 
5$^\circ$ (default criterion), 10$^\circ$, 20$^\circ$, and 30$^\circ$. 
The observations of ID1 and 5 showed a strong neutral oxygen emission line. 
This line became negligible when the ELV value was changed from 5$^\circ$ to 10$^\circ$. 
Therefore, we adapted the ELV value of 10$^\circ$. 
The exposure times of ID1 and 5 decreased by $\sim$1\% and $\sim$11\%, respectively. 
The default criterion was applied to the other observations because 
no significant neutral oxygen emission line was observed. 
The exposure times of ID1--5 are summarized in Table \ref{tab1}. 


We also checked spectra at different DYE\_ELV values of 
20$^\circ$ (default criterion), 30$^\circ$, 40$^\circ$, 50$^\circ$, and 60$^\circ$. 
The neutral oxygen emission lines of ID1 and 5 became negligible 
when the DYE\_ELV values were set to 30$^\circ$ and 40$^\circ$, respectively, 
reducing the exposure times by $\sim$14\% and $\sim$38\%. 
Considering a significant loss of the exposure time, we decided to change the ELV value rather than the DYE\_ELV value. 

\section{Light Curves}

For each observation, we plot X-ray light curves extracted from the TDX region in the 0.5--0.7 keV band 
in figures \ref{fig3}--\ref{fig7}. 
This band contains O\,\emissiontype{VII} and O\,\emissiontype{VIII} emission lines often seen in geocoronal SWCX events. 
The 0.5--0.7 keV count rate shows two features. 
One is a sudden enhancement during the observations of ID1, 3, and 5. 
The other is a gradual one during those of ID2 and 4. 
There appear to be some increases just before the sudden ones of ID1, 3, and 5. 
These temporal variations should be related to solar wind and geomagnetic events. 


Hereafter, we define the {\it stable}, {\it pre-flare}, and {\it flare} periods as indicated by 
the black bars in figures \ref{fig3}--\ref{fig7}. 
The average rates or the total counts divided by the exposure time during each period are shown in table \ref{tab2}. 
That of ID1 increased by a factor of $\sim$3 during the {\it pre-flare} period, while 
those of ID3 and 5 increased by $\sim$34\% and $\sim$20\% 
from the {\it stable} period to the {\it pre-flare} period, respectively. 
That of ID1 increased by a factor of $\sim$4 during the {\it flare} period, while 
those of ID2--5 increased by a factor of $\sim$2 during the {\it flare} period. 


We then plot X-ray light curves in the 2.5--5 keV band. 
This band is composed of a non-SWCX continuum, e.g., originating from enhanced particle backgrounds.
The 2.5--5 keV count rate shows some abrupt changes during the observations of ID1--3 but 
less variabilities during those of ID4 and 5. 
These changes were consistent with passages through regions of low geomagnetic cut-off-rigidities. 
The average rates are shown in table \ref{tab2}. 
That of ID1 increased by $\sim$23\% from the {\it stable} period to the {\it pre-flare} period and 
by a factor of $\sim$3 during the {\it flare} period. 
That of ID2 increased by a factor of $\sim$3 during the {\it flare} period, while 
that of ID3 increased by $\sim$14\% and $\sim$40\% 
from the {\it stable} period to the {\it pre-flare} and {\it flare} periods, respectively. 
Those of ID4 and 5 were almost constant during each period. 
There remain some increases, e.g., by a factor of $\sim$2 during the {\it flare} period of ID1 and 2, 
even after excluding periods where geomagnetic cut-off-rigidities are less than 8 GV, 
which is stricter than the default value of 4 GV. 
This indicates that more particles penetrate into the low-Earth orbit through the Earth's magnetosphere. 
The soft bands of ID1--3 may be affected by such particle-induced backgrounds. 
Therefore, we quantitatively checked their spectral contributions later in section 4. 


In figures \ref{fig3}--\ref{fig7}, we plot three representative solar wind parameters, 
proton density, velocity, and helium to proton ratio, and interplanetary magnetic field (IMF) components, 
$B_\mathrm{X}$, $B_\mathrm{Y}$, and $B_\mathrm{Z}$, in Geocentric Solar Magnetic (GSM) coordinates.
These data were taken from the WIND and ACE satellites 
orbiting around the Lagrangian point L$_{1}$ between the Sun and Earth and 
were shifted in time to account for solar wind propagations between the L$_{1}$ point to the near-Earth region
using the same method described in the OMNIWeb data documentation.\footnote{
$\langle$https://omniweb.gsfc.nasa.gov/html/ow\_data.html$\rangle$}
The solar wind phase front was assumed to be perpendicular to the ecliptic plane. 
The orientation of the phase front relative to the Sun--Earth line was determined from 
an intermediate geometrical consideration between co-rotation and convection. 
The estimated propagation times were consistent with those obtained from the OMNIWeb data products. 


The proton density and velocity show discontinuous changes related to arrivals of 
CME-induced interplanetary shocks during the observations of ID1, 3, and 5. 
The velocity rises much further during the second half one of ID3 
likely due to higher solar wind streams from a coronal hole. 
The density shows some increases during those of ID2 and 4, while 
the velocity increases during that of ID2 but decreases during that of ID4. 
The former is probably associated with co-rotating interaction regions. 
The latter might originate from solar wind inter-stream flows. 
The enhanced helium to proton ratio during those of ID1, 3, and 5 should be related with 
unusual element and ion abundances within CMEs (e.g., \cite{ric04}). 
The IMF of ID3 and 5 shows intense fluctuations just after the interplanetary shocks and 
smoothly rotating components of magnetic clouds during the passages of the CME itself. 
That of ID1 shows no fluctuations within the turbulent sheath but smooth magnetic fields within the magnetic cloud. 


We then plot the SYM-H index provided by the World Data Center for Geomagnetism, Kyoto, Japan,\footnote{
$\langle$http://wdc.kugi.kyoto-u.ac.jp/dstae/index.html$\rangle$.} 
which is a measure of geomagnetic disturbances at mid-latitudes, 
similar to the Dst index but with a much higher time resolution. 
Negative values indicate that a geomagnetic storm is in progress with an enhanced westward ring current around Earth 
(see \cite{kan21}). 


The SYM-H index reached less than $-$100 nT on 2005 August 24 and 2005 August 31, i.e., 
during the observations of ID1 and 2, which is classified as a major geomagnetic storm. 
These storms are associated with increased solar wind velocities and enhanced southward magnetic fields. 
That of ID3 experienced a moderate storm with minimum values of $-$50 nT on 2005 September 3--4. 
This storm is probably due to higher velocities but no southward magnetic fields within the magnetic cloud. 
No fluctuations of magnetic fields and lower velocities resulted in less deflections during that of ID4. 
There were no dramatic decreases because magnetic fields remained northward within the magnetic cloud with 
lower velocities during that of ID5. 
The compression of the magnetopause resulted in positive values and changes 
just after the interplanetary shocks of ID1, 3, and 5. 

\section{Spectrum}

We extracted spectra from the TDX region during the {\it stable}, {\it pre-flare}, and {\it flare} periods. 
These spectra include instrument and sky backgrounds. 
The former remains almost constant during each observation thanks to the low-Earth orbit of Suzaku. 
The latter consists mainly of diffuse Galactic and extragalactic emissions. 
Their spectral features do not vary temporally. 
Therefore, we assumed background components to be constant during each observation. 


Figure \ref{fig8} shows spectra produced by subtracting the {\it stable} period from the {\it pre-flare} period, 
representing enhanced components during the {\it pre-flare} period. 
These spectra contained oxygen emission lines between 0.5 keV and 0.7 keV. 
There were some emission lines constituted of highly ionized carbon and nitrogen below 0.5 keV. 
We fitted each spectrum with a theoretical model constructed by \citet{bod07}. 
This model includes cross sections for transition lines from highly charged ions (C\,\emissiontype{V}, C\,\emissiontype{VI}, 
N\,\emissiontype{VI}, N\,\emissiontype{VII}, O\,\emissiontype{VII}, and O\,\emissiontype{VIII}) 
in collision with atomic hydrogen for several velocities. 
The normalization of the principal transition with the largest cross section in each ion was considered to be a free parameter, 
while those of the other transitions were fixed according to the relative cross sections of each principal transition. 
For each spectral fitting, we adopted a collision velocity of 600 km s$^{-1}$, 
which is close to average solar wind velocities during the {\it pre-flare} period. 
We added an extra Gaussian to reproduce the lowest-energy emission line around 0.25 keV. 
The best-fit parameters are summarized in table \ref{tab3}. 


We then subtracted the {\it stable} period from the {\it flare} period. 
Figures \ref{fig9} and \ref{fig10} show the resultant spectra for individual observations. 
These spectra contained a series of emission lines from highly ionized 
carbon, nitrogen, and oxygen between 0.3 keV and 0.7 keV. 
We fitted each spectrum with the Bodewits model and one or two extra Gaussians around 0.25 keV. 
We adopted a collision velocity of 400 km s$^{-1}$ for the spectral fittings of ID2, 4, and 5 and 
that of 600 km s$^{-1}$ for those of ID1 and 3. 
The spectra of ID3 and 5 showed some excess emission lines from highly ionized 
neon, magnesium, and silicon above 0.7 keV. 
Therefore, we added 14 narrow Gaussians detected by \citet{car11}, 
representing such emission lines between 0.7 keV and 2 keV. 
The other ones showed no significant excess emission lines above 1 keV. 
The best-fit parameters are summarized in table \ref{tab4}. 


The spectra of ID1--3 have a potential influence of enhanced particle backgrounds. 
Similar particle-induced backgrounds have been observed in past geocoronal SWCX events most probably 
due to soft protons funneled by the telescope onto the detector as described in \citet{car10}. 
These particles can produce spectrally featureless signals in the wide band and throughout the entire field of view. 
Therefore, we fitted each spectrum with a power-law model. 
The spectrum during the {\it pre-flare} period of ID1 showed no significant excess components in the 1--5 keV band 
due to poor photon statistics (on an exposure time of $\sim$3 ks). 
Those during the {\it flare} period of ID1 and 2 had some excess components in the 1--5 keV band, while 
those during the {\it pre-flare} and {\it flare} periods of ID3 exceeded in the 1--5 keV and 2--5 keV band, respectively. 
The other ones showed no significant excess components in the hard band. 
The fitting results are summarized in tables \ref{tab3} and \ref{tab4}. 


The above power-law continuum showed a correlation between its spectral hardness and intensity, 
which become harder and stronger during intense geomagnetic storms. 
This supports that the enhanced particle populations in the Earth's magnetosphere 
during geomagnetic storms penetrate into an observer's line of sight through 
closed magnetic field lines (e.g., \cite{wal14}). 


We then extrapolated the above power-law continuum into the soft bands of ID1--3. 
The normalizations of O\,\emissiontype{VII} and O\,\emissiontype{VIII} emission lines 
during the {\it pre-flare} period of ID3 reduced by $\sim$2\% and $\sim$10\%, respectively, 
while those during the {\it flare} periods of ID1--3 by 
$\sim$6\% and $\sim$18\%, $\sim$3\% and $\sim$13\%, and $\sim$3\% and $\sim$4\%. 
These reductions are within 90\% confidence range listed in tables \ref{tab3} and \ref{tab4}. 
The enhanced particle backgrounds are almost negligible in the soft bands of ID1--3. 


We accurately extracted the O\,\emissiontype{VII} and O\,\emissiontype{VIII} line fluxes of 
the five geocoronal SWCX events. 
These values reached several tens of LU, where LU is photons s$^{-1}$ cm$^{-2}$ str$^{-1}$, 
which is related to solar wind and geomagnetic events, i.e., CMEs and geomagnetic storms. 

\section{Geocoronal SWCX Model}

We describe how to model geocoronal SWCX emission. 
This emission can be estimated from integral of emissivities along an observer's line of sight. 
For a single ion species, its intensity is expressed by the following equation: 
\begin{eqnarray}
I_\mathrm{SWCX} = 
\frac{1}{4\pi} \ \int \ n_\mathrm{H} \ n_\mathrm{ion} \ v_\mathrm{ion} \ \alpha \ ds, 
\label{eq1} 
\end{eqnarray}
where $n_\mathrm{H}$ is the density of the neutral hydrogen atom in the Earth's exosphere, 
$n_\mathrm{ion}$ and $v_\mathrm{ion}$ correspond to the density and the velocity of the solar wind ion species of interest, 
$\alpha$ accounts for the charge exchange cross section and transition probability for relevant emission lines, 
and $ds$ is the step length of integration. 
Below we explain these parameters. 

\subsection{Exospheric Neutral Hydrogen Distribution}

To describe neutral exospheric densities as a function of radial distances from the Earth, 
we used the simplified formula of \citet{cra01}: $n_\mathrm{H} = n_\mathrm{H0} \ (10 \ R_\mathrm{E}/r)^3$ 
with $n_\mathrm{H0} = 25 \ \mathrm{cm}^{-3}$, 
which is an approximation of results from Monte Carlo simulations 
for several values of insolation at solstice and equinox \citep{hod94}. 
The Hodges model is compatible with some measurements of hydrogen distributions using 
Lyman-$\alpha$ column brightnesses from the night side of the Earth (e.g., \cite{ost03}). 


There are several models of exospheric densities deduced from remote observations using 
Lyman-$\alpha$ (e.g., \cite{bai11,bal19}),  energetic neutral atoms (e.g., \cite{fus10}), and soft X-rays (e.g., \cite{con19}), 
showing various densities ranging roughly from 5 cm$^{-3}$ to 50 cm$^{-3}$ at 10 $R_\mathrm{E}$. 
The simplified formula we adopted is an intermediate of these different models. 
We tested each model and found that these uncertainties can change line intensities by a factor of $\sim$2--3. 

\subsection{Solar Wind Ion Data}

The O\,\emissiontype{VII} emission lines are produced by O$^{7+}$ ions 
undergoing charge exchange to become O$^{6+}$ ions in excited states, 
while the O\,\emissiontype{VIII} emission lines by O$^{8+}$ ions. 
These ion densities can be deduced from that of solar wind proton multiplied by 
helium to proton ratio, oxygen to helium ratio, and oxygen charge state fraction of interest: 
\begin{eqnarray}
n_\mathrm{O^{q+}} = n_\mathrm{p} \ 
\left[\frac{\mathrm{He}}{\mathrm{p}}\right] \ 
\left[\frac{\mathrm{O}}{\mathrm{He}}\right] \ 
\left[\frac{\mathrm{O^{q+}}}{\mathrm{O}}\right], 
\label{eq2} 
\end{eqnarray}
where the proton density can be obtained from the OMNIWeb data products where cross-calibration issues 
between WIND/SWE and ACE/SWEPAM have already been handled. 


The He/p ratio can be taken from WIND/SWE and ACE/SWEPAM. 
The ACE/SWEPAM He/p ratio suffers from a significant problem 
(see ``Important Notes'' of the OMNIWeb data documentation).\footnotemark[6] 
The WIND/SWE He/p ratio may be more appropriate. 


The other parameters are only available from ACE/SWICS. 
We used SWICS 1.1 level 2 version 4.09 data processed on 2015 June 8. 
The instrument team of ACE/SWICS estimated uncertainties of 30\% for most parameters (see release notes).\footnote{
$\langle$https://izw1.caltech.edu/ACE/ASC/DATA/level2/ssv4/swics\_lv2\_v4\_release\_notes.txt$\rangle$}
Below we describe solar wind ion data during each observation. 


\begin{itemize}

\item 
ID1: The WIND/SWE He/p ratio was sparse during most of the observation except for a part of the {\it stable} period. 
The ACE/SWEPAM He/p ratio lacked during the {\it flare} period. 
Therefore, we used the ACE/SWICS He/p ratio during the entire observation. 
The ACE/SWICS O/He ratio and oxygen charge state fractions were available. 

\item 
ID2: The WIND/SWE He/p ratio lacked during a part of the {\it stable} period. 
Therefore, we used the ACE/SWEPAM He/p ratio during a part of the {\it stable} period. 
The ACE/SWICS He/p ratio was used during periods where the ACE/SWEPAM He/p ratio was not available. 
The ACE/SWICS O/He ratio and oxygen charge state fractions were available. 

\item 
ID3: The WIND/SWE He/p ratio was sparse during most of the observation except for the {\it flare} period. 
Therefore, we used the ACE/SWEPAM He/p ratio during most of the observation except for the {\it flare} period. 
The ACE/SWICS He/p ratio was used during periods where the ACE/SWEPAM He/p ratio was not available. 
The ACE/SWICS O/He ratio and oxygen charge state fractions were available. 

\item 
ID4: The WIND/SWE He/p ratio was available during the entire observation. 
The ACE/SWICS O/He ratio and oxygen charge state fractions were available. 

\item 
ID5: The WIND/SWE He/p ratio was sparse during the {\it flare} period. 
Therefore, we used the ACE SWEPAM He/p ratio during the {\it flare} period. 
The ACE/SWICS O/He ratio and oxygen charge state fractions were not available 
after hardware anomalies altered instrumental operational states on 2011 August 23. 
We thus refer to slow (442 km s$^{-1}$) and fast (810 km s$^{-1}$) solar wind ion abundances listed in \citet{sch00}. 
The solar wind velocity ranged roughly from 400 km s$^{-1}$ to 500 km s$^{-1}$ during the entire observation. 
Therefore, we used the slow solar wind values. 
The He/O ratio was assumed to be 78. 
The O$^{7+}$/O and O$^{8+}$/O ratios were set to 0.20 and 0.07, respectively. 

\end{itemize}


The remaining parameter is oxygen ion velocities. 
These ion velocities were assumed to be the same as the proton values. 
Thermal velocities were added in quadrature to them: 
\begin{eqnarray}
v_\mathrm{O^{q+}} \simeq \sqrt{v_\mathrm{p}^2 + \frac{3 k_\mathrm{B} T_\mathrm{p}}{m_\mathrm{p}}}, 
\label{eq3} 
\end{eqnarray}
where $k_\mathrm{B}$ is the Boltzmann constant, 
$T_\mathrm{p}$ is the proton temperature, and 
$m_\mathrm{p}$ is the proton mass. 
The proton velocity and temperature can be obtained from the OMNIWeb data products. 

\subsection{Charge Exchange Cross Section}

Charge exchange cross sections and line yields for each transition of 
O$^{7+}$ (seven transitions) and O$^{8+}$ (five transitions) ions were taken from \citet{bod07}. 
To obtain values corresponding to a particular collision velocity, we interpolated the tabled values for the five ones 
(200, 400, 600, 800, and 1000 km s$^{-1}$). 
\citet{bod07} estimated uncertainties to be approximately 20\%.

\subsection{Line of Sight Integration}

Magnetosheath plasma populations are responsible for soft X-ray emitters, while 
magnetospheric ones contain few highly charged ions required to produce soft X-rays. 
Therefore, we assumed relevant ion densities to be zero inside the magnetopause and outside the bow shock. 
The magnetopause and bow shock positions were determined from empirical models of 
\citet{shu98} and \citet{mer05}, respectively. 
The former is parameterized by the IMF $B_\mathrm{Z}$ component in GSM coordinates and 
the solar wind dynamic pressure, while the latter by the upstream Alfv\'enic Mach number. 
Both models represent average positions for particular solar wind parameters so that abrupt changes, 
e.g., the interplanetary shock-induced impulses of ID1, 3, and 5, are not easy to be considered. 


The magnetospheric cusp is a narrow throat of magnetic field lines poleward of the last closed field line 
on the day side of the Earth. 
These magnetic field lines are open and allow solar wind plasma to enter deep into the near-Earth region 
with higher exospheric densities (e.g., \cite{wal16}). 
The magnetopause model does not take into account cusp geometries. 
Therefore, we used the Earth's magnetic field model (\cite{tsy05} and references therein) 
to examine whether magnetic field lines along line of sights are closed, 
open but connected to the north or south poles, or not connected to Earth. 
The Tsyganenko model is a semi-empirical best-fit representation for the Earth's magnetic field. 
Its input parameters are the solar wind dynamic pressure, the Dst index, 
the IMF $B_\mathrm{Y}$ and $B_\mathrm{Z}$ components in GSM coordinates, 
and a set of variable weight coefficients provided by TS05 web repository.\footnote{
$\langle$http://geo.phys.spbu.ru/\~{t}syganenko/TS05\_data\_and\_stuff/$\rangle$}
We traced magnetic field lines along an observer's line of sight and determined points 
whose magnetic fields are open to the day side of the Earth, i.e., 
cusp regions where solar wind plasma can exist. 


Figure \ref{fig11} shows configuration of our modeled magnetopause, bow shock, and magnetic field lines 
during the observation of ID3. 
We thus integrate emissivities at magnetosheath and cusp regions along an observer's line of sight. 
The solar wind plasma is being shocked downstream of the bow shock. 
We considered this effect using the Rankine--Hugoniot equations. 
The polytropic index was set to 1.46 (e.g., \cite{tot95}). 
The shocked solar wind plasma was assumed to be uniformly distributed at each integral point 
along an observer's line of sight. 


Provided that magnetic field models have little uncertainties, our model uncertainties are dominated 
mainly by an exospheric neutral hydrogen distribution model 
and then by solar wind ion data and charge exchange cross sections. 
The summed uncertainty becomes a factor of $\sim$3--5. 

\section{Results}

\subsection{Average Line Flux}

To examine our model accuracy, we estimated average line fluxes during each observation. 
Table \ref{tab5} gives modeled intensities of O\,\emissiontype{VII} and O\,\emissiontype{VIII} emission lines 
during the {\it stable}, {\it pre-flare}, and {\it flare} periods. 
For comparison with the observed values, we subtracted the modeled line fluxes during the {\it stable} period from 
those during the {\it pre-flare} and {\it flare} periods. 
Figure \ref{fig12} shows ratios between the observed and modeled intensities 
for the O\,\emissiontype{VII} and O\,\emissiontype{VIII} line fluxes during each observation. 


For the O\,\emissiontype{VII} line flux, considering our model uncertainties of a factor of $\sim$3--5, 
we found that the model reproduced the data except for the result of ID2. 
This indicates that our model is useful for estimating the contribution of the O\,\emissiontype{VII} emission line. 
The observation of ID2 may be affected by solar wind injections into the inner magnetosphere 
during intense geomagnetic storms. 
\citet{ebi09} suggested that high-charge state oxygen ions were transported to the inner magnetosphere 
from the night side of the high-latitude magnetopause during intense geomagnetic storms. 
This situation seems to be consistent with our Suzaku observation whose line-of-sight direction was toward 
the night side of the high-latitude magnetosheath during a major geomagnetic storm reaching less than $-$100 nT. 
Not only O$^{7+}$ ions in the magnetosheath but also in the inner magnetosphere may be responsible for soft X-ray emitters. 


The O\,\emissiontype{VIII} line flux was more underestimated compared to the O\,\emissiontype{VII} line flux. 
Even considering our model uncertainties, such large discrepancies can not be explained. 
This suggests that further uncertainties exist in solar wind ion abundances, i.e., 
the measurement accuracy of O$^{8+}$ ions may be worse than that of O$^{7+}$ ions 
due to poor counting statistics. 
Figure \ref{fig13} shows comparison between the O$^{8+}$/O$^{7+}$ ion ratio measured by ACE/SWICS 
and the O\,\emissiontype{VIII}/O\,\emissiontype{VII} flux ratio deduced from geocoronal SWCX spectra 
during each observation. 
The ratios of the Suzaku spectra tend to be larger than those of ACE/SWICS. 
This supports that O$^{8+}$ ions may be poorly measured by ACE/SWICS. 


To check a possible effect due to our simplified magnetosheath model, we tested an MHD model for the observation of ID4. 
Most MHD simulations have some difficulties handling the near-cusp region, e.g., within 3 $R_\mathrm{E}$. 
This observation whose line-of-sight direction was toward the flanks of the magnetosheath is a good example to 
verify our model. 
We adopted Block-Adaptive-Tree Solar Wind Roe-Type Upwind Scheme \citep{tot05,tot12}, 
available via the Community Coordinated Modeling Center facility through their public Runs on Request system,\footnote{
$\langle$http://ccmc.gsfc.nasa.gov$\rangle$.} 
and downloaded an output file that had been run for other projects relevant to our Suzaku observation 
whose run name is ``David\_Sibeck\_123011\_1.'' 
This output covers from 12:00 UT on 2005 October 29 to 12:00 UT on 2005 October 30, 
corresponding roughly to the last half of the {\it stable} period and the first half of the {\it flare} period. 
The O\,\emissiontype{VII} line fluxes during the above periods were estimated to be 0.50 LU and 4.14 LU. 
These values are $\sim$1.4 and $\sim$2 times smaller than those obtained from our model 
during the corresponding periods, i.e., 0.72 LU and 8.10 LU, respectively. 
The O\,\emissiontype{VIII} line fluxes were 0.027 LU and 0.008 LU, which are $\sim$1.1 and $\sim$1.5 times smaller than 
our model values, i.e., 0.030 LU and 0.012 LU, respectively. 
The O\,\emissiontype{VII} line flux becomes more consistent with the observed value, 
while the O\,\emissiontype{VIII} line flux remains more underestimated. 


To estimate more accurate solar wind ion fluxes for the observation of ID5, we tested an empirical equation of \citet{kaa20}. 
This equation is based on the analysis of the stacked data before the anomalies of ACE/SWICS. 
The O$^{7+}$ and O$^{8+}$ ion fluxes can be deduced from the O$^{7+}$/O$^{6+}$ ratio. 
The O\,\emissiontype{VII} line fluxes during the {\it stable}, {\it pre-flare}, and {\it flare} periods were estimated to be 
0.60 LU, 1.60 LU, and 2.87 LU, respectively. 
These values are $\sim$3, $\sim$7, and $\sim$5 times smaller than our model values. 
Those during the {\it pre-flare} and {\it flare} periods subtracted by that during the {\it stable} period were 
1.00 LU and 2.28 LU, respectively, which agree with the observed values within a factor of $\sim$6 and $\sim$3. 
These differences become more significant than those obtained from our model. 
The O\,\emissiontype{VIII} line fluxes during the {\it stable}, {\it pre-flare}, and {\it flare} periods were 
0.010 LU, 0.080 LU, and 0.429 LU, respectively, which are 
$\sim$35, $\sim$36, and $\sim$6 times smaller than our model values. 
Those during the {\it pre-flare} and {\it flare} periods subtracted by that during the {\it stable} period were 
0.070 LU and 0.419 LU, respectively, which are $\sim$83 and $\sim$82 times smaller than the observed values. 
There remain more significant differences compared with our model results. 


The remaining concern is a potential contribution from heliospheric SWCX emission. 
The line-of-sight direction of Suzaku becomes parallel to the local Parker spiral, 
i.e., the orientation of the phase front near the Earth, and has a long pass length of $\sim$1 AU, 
thereby producing potentially significant heliospheric SWCX contributions 
with time scales similar to geocoronal SWCX emission. 
\citet{kun15} and \citet{kun19} provided maps of temporal variabilities from heliospheric SWCX emission and 
their correlations with local solar wind fluxes near equator regions. 
The line-of-sight directions of ID1 and 4 were toward the near-equator region whose 
temporal variability and correlation are relatively high but not strong. 
Using the Parker spiral equations \citep{pak58}, we estimated the pass length of 
ID1 and 4 to be $\sim$0.4 AU and $\sim$0.2 AU, respectively.
The line-of-sight directions of ID2, 3, and 5 were toward the near-polar region whose path length is $\sim$0.4 AU.
The O\,\emissiontype{VII} and O\,\emissiontype{VIII} line fluxes were estimated to be 1.09 LU and 0.60 LU, respectively, 
assuming slow solar wind parameters and cross sections for hydrogen and helium atoms (\cite{kou06} and references therein), 
atomic hydrogen and helium densities of 0.001 cm$^{-3}$ and 0.005 cm$^{-3}$ near the Earth (e.g., \cite{cra01}), 
and a path length of 0.2 AU. 
Those were 0.19 LU and 0.00 LU for fast solar wind values.
The local Parker spiral contributions for O\,\emissiontype{VII} and O\,\emissiontype{VIII} emission lines were 
estimated to be 0.90 LU and 0.60 LU, respectively, considering the difference between 
uniformly high and low emissivities originating from slow and fast solar winds. 
The O\,\emissiontype{VII} line flux is $\sim$4--46 times smaller than the observed values during the observations of ID1--5, 
while the O\,\emissiontype{VIII} line flux is $\sim$0.7--57 times smaller.
Although the accurate estimate of the local Parker spiral contribution needs more accurate 
solar wind propagations and interstellar neutral distributions, 
the bright SWCX events we analyzed are dominated mainly by geocoronal SWCX emission. 

\subsection{Model Light Curve} 

We simulated O\,\emissiontype{VII} and O\,\emissiontype{VIII} light curves in units of LU. 
Figures \ref{fig14}--\ref{fig18} show results of ID1--5. 
We plot X-ray light curves extracted from the TDX region in the 0.52--0.6 and 0.6--0.7 keV band 
along with solar wind proton flux, oxygen to proton ratio, and oxygen ion fractions. 
The count rate was converted into the line flux per solid angle using the area of the TDX region and 
the spectral fitting result during the {\it flare} period. 
The modeled light curves with a time bin of 256 s were binned into the same bin of the observed ones 
and scaled by the ratio between the observed and modeled intensities during the {\it flare} period. 
The background flux was estimated from the average rate during the {\it stable} period. 


The above scaled model reproduced the observed temporal variations in the 0.52--0.6 and 0.6--0.7 keV band 
except for those during the {\it pre-flare} period of ID5. 
This indicates that our model is capable for predicting the O\,\emissiontype{VII} and O\,\emissiontype{VIII} light curves. 
The discrepancies of ID5 are most probably due to constant oxygen ion fractions. 
The time-variable oxygen ion fluxes deduced from an empirical equation improve such discrepancies, 
while its scaling factors become worse as mentioned former in subsection 7.1. 


In figures \ref{fig19}--\ref{fig23}, we plot enlarged views with shorter time bins during the {\it pre-flare} and {\it flare} period. 
There are some spike bins due to line-of-sight directions passing through the near-cusp region, 
e.g., within 3 $R_\mathrm{E}$, during the orbital motion of Suzaku. 
These spikes were reproduced by our scaled model. 
This supports that the strongest emitters are present in the magnetospheric cusps and 
geocoronal SWCX emission is useful for capturing cusp geometries and motions. 
Below we describe the result for each observation. 


\begin{itemize}

\item 
ID1: The O\,\emissiontype{VII} and O\,\emissiontype{VIII} light curves vary due to 
increased solar wind proton flux and oxygen ion fractions. 
The oxygen to proton ratio increases but has less contributions to the observed temporal variations. 
There are some spike bins due to the line of sight direction passing through the southern polar cusp. 

\item 
ID2: The O\,\emissiontype{VII} and O\,\emissiontype{VIII} light curves vary due to 
increased solar wind proton flux and oxygen to proton ratio. 
The oxygen ion fractions decrease during the {\it flare} period. 
There are some spike bins due to the line of sight direction passing through the southern polar cusp. 

\item 
ID3: The O\,\emissiontype{VII} and O\,\emissiontype{VIII} light curves vary due to 
increased solar wind proton flux and oxygen to proton ratio. 
The O$^{7+}$/O ratio increases and has further contributions to the observed temporal variations during the {\it flare} period. 
The O$^{8+}$/O ratio has less variabilities during the entire observation. 
There are a lot of spike bins due to the line of sight direction passing through the northern polar cusp. 

\item 
ID4: The O\,\emissiontype{VII} light curve varies due to increased solar wind proton flux and oxygen to proton ratio. 
The O$^{7+}$/O ratio increased but the O$^{8+}$/O decreased during the {\it flare} period. 
There are no spike bins because the line-of-sight direction was toward the flanks of the magnetosheath. 

\item 
ID5: The O\,\emissiontype{VII} and O\,\emissiontype{VIII} light curves vary due to 
increased solar wind proton flux and oxygen to proton ratio. 
The oxygen ion fractions have some contributions to the observed temporal variations 
during the {\it pre-flare} and {\it flare} periods. 
There are a lot of spike bins due to the line of sight direction passing through the southern polar cusp. 

\end{itemize}


We need more data to calibrate our model and to reduce uncertainties 
problematic for astronomical observations. 
Future high-resolution and high-sensitivity X-ray spectroscopy missions such as 
XRISM\footnote{$\langle$https://xrism.isas.jaxa.jp$\rangle$} and 
Athena\footnote{$\langle$https://www.cosmos.esa.int/web/athena$\rangle$} 
will provide us with more SWCX events and more detailed information such as solar wind compositions, kinematics, 
and charge exchange processes as demonstrated in the X-ray micro-calorimeter instrument onboard Hitomi (e.g, \cite{ezo21}). 
On the other hand, geocoronal SWCX emission is suggested to be used to X-ray imaging of the Earth's magnetosphere 
as planned in future missions such as SMILE \citep{bra18} and GEO-X \citep{ezo18}. 

\section{Summary}

In this paper, we have built the empirical model to predict time-variable geocoronal SWCX emission 
and have examined the model in the five Suzaku observations of the bright geocoronal SWCX events. 
We re-analyzed the Suzaku data so that line intensities of geocoronal SWCX emission are accurately extracted 
in the same manner. 
The method established by \citet{ezo11} was employed. 
For model comparison, we focused on the strong O\,\emissiontype{VII} and O\,\emissiontype{VIII} emission lines 
seen in the 0.5--0.7 keV band. 


In the modeling, we took into account time-variable solar wind ion fluxes and abundances using the WIND and ACE data. 
To describe exospheric neutral hydrogen distributions, we adopted a simple formula built by \citet{cra01}. 
Charge exchange cross sections were taken from values shown in \citet{bod07}, 
which are based on ground experiments and theoretical predictions. 
The magnetopause and bow shock positions were determined from the empirical models of 
\citet{shu98} and \citet{mer05}, respectively. 
To consider the line-of-sight direction passing through the near-cusp region during the orbital motion of Suzaku, 
we traced the magnetic field lines along the observer's line of sight using the magnetic field model of \citet{tsy05}. 
The decelerated and heated solar wind plasma downstream of the bow shock was represented by 
the Rankine--Hugoniot equations. 


Using the model, we estimated the O\,\emissiontype{VII} line flux and found that 
the model agreed with the data except for one case in which 
the line-of-sight direction was toward the night side of the high-latitude magnetosheath and 
the major geomagnetic storm was observed. 
The solar wind injection into the inner magnetosphere may contribute to geocoronal SWCX emission. 
The O\,\emissiontype{VIII} line flux was not consistent with the data in all the five cases. 
These discrepancies can not be explained even considering possible model uncertainties. 
This suggests that further uncertainties exist in the solar wind ion data concerning highly stripped ion states. 
We simulated geocoronal SWCX light curves and found that the modeled light curves after scaling are consistent with 
the data including some spike behaviors due to the line-of-sight direction passing through the near-cusp regions 
associated with the low-Earth orbit. 


Although more SWCX events are needed to examine such tendencies, 
this model can provide a new estimation of geocoronal SWCX emission including light curves 
for future X-ray astronomy missions as well as X-ray imaging missions of the Earth's magnetosphere. 


\begin{table*}[t]
\tbl{Suzaku observation log of five bright geocoronal SWCX events.}{
\begin{tabular}{ccccccc} \hline
ID & ObsID & Date & Target name & Target coordinates & Effective exposure & TDX area \\ 
& & & & (RA, Dec)$_\mathrm{J2000.0}$ & (ks) & (arcmin$^2$) \\ \hline
1\footnotemark[$*$] & 100009010 & 2005 August 23--24 & PSR B1509$-$58 & (\timeform{228D.4837}, \timeform{-59D.1356}) & 60.7 & 44.4 \\ 
2\footnotemark[$\dagger$] & 100014010 & 2005 August 31 & 1E0102.2$-$7219 & (\timeform{16D.0100}, \timeform{-72D.0333})& 24.3 & 71.0 \\ 
3\footnotemark[$\ddagger$] & 100018010 & 2005 September 2--4 & North Ecliptic Pole & (\timeform{272D.8000}, \timeform{66D.0000}) & 106.2 & 227.3 \\ 
4\footnotemark[$\S$] & 500009010 & 2005 October 28--30 & Galactic Ridge & (\timeform{281D.0000}, \timeform{-4D.0700}) & 93.4 & 198.8 \\ 
5\footnotemark[$\|$] & 508072010 & 2013 April 11--15 & 0509$-$67.5 & (\timeform{77D.3927}, \timeform{-67D.5253}) & 157.4 & 174.7 \\ \hline
\end{tabular}} 
\label{tab1}
\begin{tabnote}
\footnotemark[$*$] \citet{ezo11}, 
\footnotemark[$\dagger$] \citet{ish13}, 
\footnotemark[$\ddagger$] \citet{fuj07}, 
\footnotemark[$\S$] \citet{ezo10}, and 
\footnotemark[$\|$] \citet{ish19}. 
\end{tabnote}
\end{table*}

\begin{table*}[t]
\tbl{XIS 1 count rates extracted from the TDX region.\footnotemark[$*$]}{
\begin{tabular}{ccccc} \hline
ID & Energy band & {\it Stable} & {\it Pre-flare} & {\it Flare} \\ \hline
1 & 0.5--0.7 keV & 0.12 $\pm$ 0.01 & 0.34 $\pm$ 0.05 & 0.53 $\pm$ 0.02 \\
& 2.5--5 keV & 0.47 $\pm$ 0.02 & 0.57 $\pm$ 0.07 & 1.39 $\pm$ 0.04 \\ \hline
2 & 0.5--0.7 keV & 0.31 $\pm$ 0.02 & -- & 0.67 $\pm$ 0.03 \\
& 2.5--5 keV & 0.17 $\pm$ 0.02 & -- &0.52 $\pm$ 0.02 \\ \hline
3 & 0.5--0.7 keV & 0.20 $\pm$ 0.01 & 0.27 $\pm$ 0.01 & 0.36 $\pm$ 0.01 \\
& 2.5--5 keV & 0.19 $\pm$ 0.01 & 0.22 $\pm$ 0.01 & 0.27 $\pm$ 0.01 \\ \hline
4 & 0.5--0.7 keV & 0.07 $\pm$ 0.01 & -- & 0.12 $\pm$ 0.01 \\
& 2.5--5 keV & 0.37 $\pm$ 0.01 & -- & 0.39 $\pm$ 0.01 \\ \hline
5 & 0.5--0.7 keV & 0.17 $\pm$ 0.01 & 0.20 $\pm$ 0.01 & 0.33 $\pm$ 0.01 \\
& 2.5--5 keV & 0.12 $\pm$ 0.01 & 0.11 $\pm$ 0.01 & 0.12 $\pm$ 0.01 \\ \hline
\end{tabular}} 
\begin{tabnote}
\footnotemark[$*$] In units of 10$^{-3}$ counts s$^{-1}$ arcmin$^{-2}$. 
\end{tabnote}
\label{tab2}
\end{table*}

\begin{table*}[t]
\tbl{Best-fitting parameters of the spectra shown in figure \ref{fig8}.\footnotemark[$*$]}{
\begin{tabular}{ccccc} \hline
& & ID1 & ID3 & ID5 \\ \hline 
C band lines & $E_c$\footnotemark[$\dagger$] & -- & 236 $^{+16}_{-49}$ & 283 $^{+2}_{-21}$ \\ 
& Norm.\footnotemark[$\ddagger$] & -- & 4.7 $\pm$ 4.5 & 5.9 $^{+3.4}_{-3.5}$ \\ \hline 
C\,\emissiontype{V} (299 eV) & Norm.\footnotemark[$\ddagger$] & 128 $\pm$ 106 & 28 $\pm$ 20 & -- \\ 
C\,\emissiontype{VI} (367 eV) & Norm.\footnotemark[$\ddagger$] & 7.8 ($<25.2$) & 2.2 ($<5.5$) & -- \\ 
N\,\emissiontype{VI} (420 eV) & Norm.\footnotemark[$\ddagger$] & 3.9 ($<13.7$) & 2.9 $\pm$ 2.1 & -- \\ 
N\,\emissiontype{VII} (500 eV) & Norm.\footnotemark[$\ddagger$] & 19 $\pm$ 11 & 0.56 ($<2.03$) & -- \\ 
O\,\emissiontype{VII} (561 eV) & Norm.\footnotemark[$\ddagger$] & 11 ($<23$) & 5.9 $\pm$ 2.1 & 5.6 $\pm$ 3.9 \\ 
O\,\emissiontype{VIII} (653 eV) & Norm.\footnotemark[$\ddagger$] & 4.2 ($<12.0$) & 1.5 $\pm$ 1.3 & 5.8 $\pm$ 2.3 \\ \hline 
$\chi^2$/d.o.f & & 10.32/11 & 24.62/25 & 7.97/14 \\ \hline 
Power-law & Photon index $\Gamma$ & -- & 0.35 $^{+0.68}_{-0.35}$ & -- \\ 
& Norm.\footnotemark[$\S$] & -- & 1.0 $\pm$ 0.5 & -- \\ \hline 
$\chi^2$/d.o.f & & -- & 22.46/33 & -- \\ \hline
\end{tabular}}
\label{tab3}
\begin{tabnote}
\footnotemark[$*$] All the line widths are fixed at 0 eV. \\
\footnotemark[$\dagger$] $E_c$ is the line center energy in units of eV. \\
\footnotemark[$\ddagger$] Normalization in units of photons s$^{-1}$ cm$^{-2}$ str$^{-1}$. \\
\footnotemark[$\S$] Normalization in units of photons s$^{-1}$ cm$^{-2}$ str$^{-1}$ keV$^{-1}$ at 1 keV. 
\end{tabnote}
\end{table*}

\begin{table*}[t]
\tbl{Best-fitting parameters of the spectra shown in figures \ref{fig9} and \ref{fig10}.\footnotemark[$*$]}{
\begin{tabular}{ccccccc} \hline
& & ID1 & ID2 & ID3 & ID4 & ID5 \\ \hline 
C band lines & $E_c$ &  244 $^{+6}_{-8}$ & 216 $^{+8}_{-14}$ & 213 $^{+4}_{-6}$ & 242 $^{+14}_{-20}$ & 246 $^{+6}_{-8}$ \\
& Norm. & 44 $\pm$ 10 & 40 $\pm$ 19 & 20 $\pm$ 6 & 5.0 $^{+2.7}_{-2.8}$ & 38 $\pm$ 11 \\
C band lines & $E_c$ & -- & 271 $^{+6}_{-11}$ & 268 $^{+5}_{-6}$ & -- & -- \\
& Norm. & -- & 20 $\pm$ 8 & 7.8 $\pm$ 2.3 & -- & -- \\ \hline 
C\,\emissiontype{V} (299 eV) & Norm. & 191 $\pm$ 47 & 55 $^{+50}_{-52}$ & 40 $^{+16}_{-17}$ & 30 $\pm$ 18 & --  \\ 
C\,\emissiontype{VI} (367 eV) & Norm. & 35 $^{+8}_{-9}$ & 38 $\pm$ 9 & 12 $\pm$ 3 & 0.89 ($<3.32$) & 25 $\pm$ 8 \\ 
N\,\emissiontype{VI} (420 eV) & Norm. & 11 $\pm$ 5 & -- & 0.12 ($<1.69$) & 2.1 $\pm$ 1.5 & -- \\ 
N\,\emissiontype{VII} (500 eV) & Norm. & 13 $\pm$ 4 & 6.7 $\pm$ 4.6 & 2.5 $\pm$ 1.2 & 1.8 $\pm$ 0.9 & 2.9 ($<6.5$) \\
O\,\emissiontype{VII} (561 eV) & Norm. & 41 $^{+5}_{-6}$ & 33 $\pm$ 6 & 7.4 $\pm$ 1.6 & 4.0 $^{+1.2}_{-1.1}$ & 6.7 $\pm$ 5.2 \\
O\,\emissiontype{VIII} (653 eV) & Norm. & 15 $^{+4}_{-3}$ & 11 $\pm$ 5 & 7.6 $\pm$ 1.2 & 0.41 ($<1.04$) & 34 $\pm$ 4 \\ \hline 
Fe\,\emissiontype{XVII} (730 eV) & Norm. & -- & -- & 1.4 $\pm$ 0.6 & -- & 2.9 $\pm$ 1.9 \\
Fe\,\emissiontype{XVII} (820 eV) & Norm. & -- & -- & 0.55 $\pm$ 0.49 & -- & -- \\
Fe\,\emissiontype{XVIII} (870 eV) & Norm. & -- & -- & -- & -- & -- \\
Ne\,\emissiontype{IX} (920 eV) & Norm. & -- & -- & 1.0 $\pm$ 0.4 & -- & 4.5 $\pm$ 1.2 \\
Fe\,\emissiontype{XX} (960 eV) & Norm. & -- & -- & -- & -- & -- \\
Ne\,\emissiontype{X} (1022 eV) & Norm. & -- & -- & 1.3 $\pm$ 0.4 & -- & 5.2 $\pm$ 0.9 \\
Ne\,\emissiontype{IX} (1100 eV) & Norm. & -- & -- & 0.37 $\pm$ 0.30 & -- & 0.62 $\pm$ 0.60 \\
Ne\,\emissiontype{X} (1220 eV) & Norm. & -- & -- & 0.10 ($<0.35$) & -- & 0.18 ($<0.67$) \\
Mg\,\emissiontype{XI} (1330 eV) & Norm. & -- & -- & 1.1 $\pm$ 0.3  & -- & 5.1 $\pm$ 0.7 \\
Mg\,\emissiontype{XII} (1470 eV) & Norm. & -- & -- & 0.64 $\pm$ 0.29 & -- & 1.4 $\pm$ 0.5 \\
Mg\,\emissiontype{XI} (1600 eV) & Norm. & -- & -- & 0.35 $\pm$ 0.29 & -- & 0.55 $\pm$ 0.42 \\
Al\,\emissiontype{XIII} (1730 eV) & Norm. & -- & -- & 0.22 ($<0.52$) & -- & 0.22 ($<0.64$) \\
Si\,\emissiontype{XIII} (1850 eV) & Norm. & -- & -- & 0.39 $\pm$ 0.27 & -- & 1.4 $\pm$ 0.5 \\ 
Si\,\emissiontype{XIV} (2000 eV) & Norm. & -- & -- & 0.31 ($<0.66$) & -- & 1.2 $\pm$ 1.1 \\ \hline
$\chi^2$/d.o.f & & 77.95/44 & 23.17/27 & 49.22/43 & 27.86/22 & 77.41/66 \\ \hline
Power-law & Photon index $\Gamma$ & $-$0.20 $\pm$ 0.05 & 0.04 $^{+0.10}_{-0.09} $ & 0.19 $^{+0.16}_{-0.13}$ & -- & -- \\ 
& Norm. & 25 $\pm$ 2 & 10 $\pm$ 1 & 2.4 $\pm$ 0.4 & -- & -- \\ \hline 
$\chi^2$/d.o.f & & 161.82/157 & 39.92/32 & 27.33/19 & -- & -- \\ \hline
\end{tabular}}
\label{tab4}
\begin{tabnote}
\footnotemark[$*$] Definitions of parameters are the same as in table \ref{tab3}. \\
\end{tabnote}
\end{table*}

\begin{table*}[t]
\tbl{Model prediction of average line fluxes.\footnotemark[$*$]}{ 
\begin{tabular}{ccccccc} \hline
ID & Emission line & {\it Stable} & {\it Pre-flare} & {\it Pre-flare} $-$ {\it Stable}\footnotemark[$\dagger$] & {\it Flare} & {\it Flare} $-$ {\it Stable}\footnotemark[$\dagger$] \\ \hline
1 & O\,\emissiontype{VII} & 0.22 & 9.52 & 9.30 & 22.53 & 22.31 \\ 
& & & & (11) & & (41) \\ 
& O\,\emissiontype{VIII} & 0.005 & 0.831 & 0.826 & 3.709 & 3.704 \\ 
& & & & (4.2) & & (15) \\ \hline
2 & O\,\emissiontype{VII} & 0.87 & -- & -- & 6.88 & 6.01 \\ 
& & & & (--) & & (33) \\ 
& O\,\emissiontype{VIII} & 0.023 & -- & -- & 0.088 & 0.065 \\ 
& & & & (--) & & (11) \\ \hline
3 & O\,\emissiontype{VII} & 0.16 & 3.26 & 3.10 & 5.91 & 5.75 \\ 
& & & & (5.9) & & (7.4) \\ 
& O\,\emissiontype{VIII} & 0.006 & 0.159 & 0.153 & 0.268 & 0.261 \\ 
& & & & (1.5) & & (7.6) \\ \hline
4 & O\,\emissiontype{VII} & 0.47 & -- & -- & 7.42 & 6.95 \\ 
& & & & (--) & & (4.0) \\ 
& O\,\emissiontype{VIII} & 0.011 & -- & -- & 0.014 & 0.003 \\ 
& & & & (--) & & (0.41) \\ \hline
5 & O\,\emissiontype{VII} & 1.73 & 11.46 & 9.73 & 13.62 & 11.89 \\ 
& & & & (5.6) & & (6.7) \\ 
& O\,\emissiontype{VIII} & 0.35 & 2.86 & 2.50 & 2.65 & 2.30 \\ 
& & & & (5.8) & & (34) \\ \hline
\end{tabular}} 
\label{tab5}
\begin{tabnote}
\footnotemark[$*$] In units of photons s$^{-1}$ cm$^{-2}$ str$^{-1}$. \\
\footnotemark[$\dagger$] Observed values are shown in parentheses. 
\end{tabnote}
\end{table*}


\begin{figure*}[t]
\begin{center}
\includegraphics[width=\textwidth]{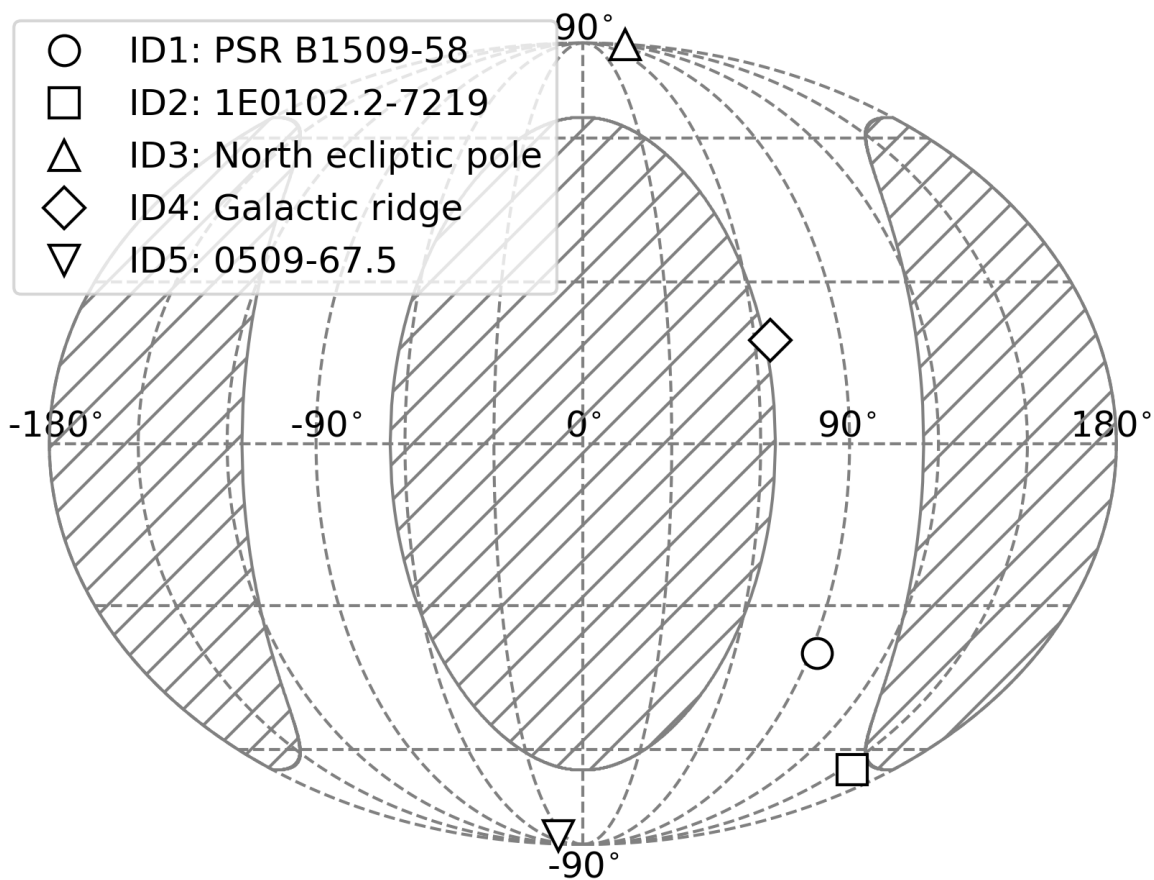} 
\end{center}
\caption{
Average line-of-sight directions in the GSE coordinate system during the observations of ID1--5. 
The hatched regions mark the Sun angle range prohibited in Suzaku observations outside 65$^\circ$--115$^\circ$. 
}
\label{fig1} 
\end{figure*}

\begin{figure*}[t]
\begin{center}
\includegraphics[width=\textwidth]{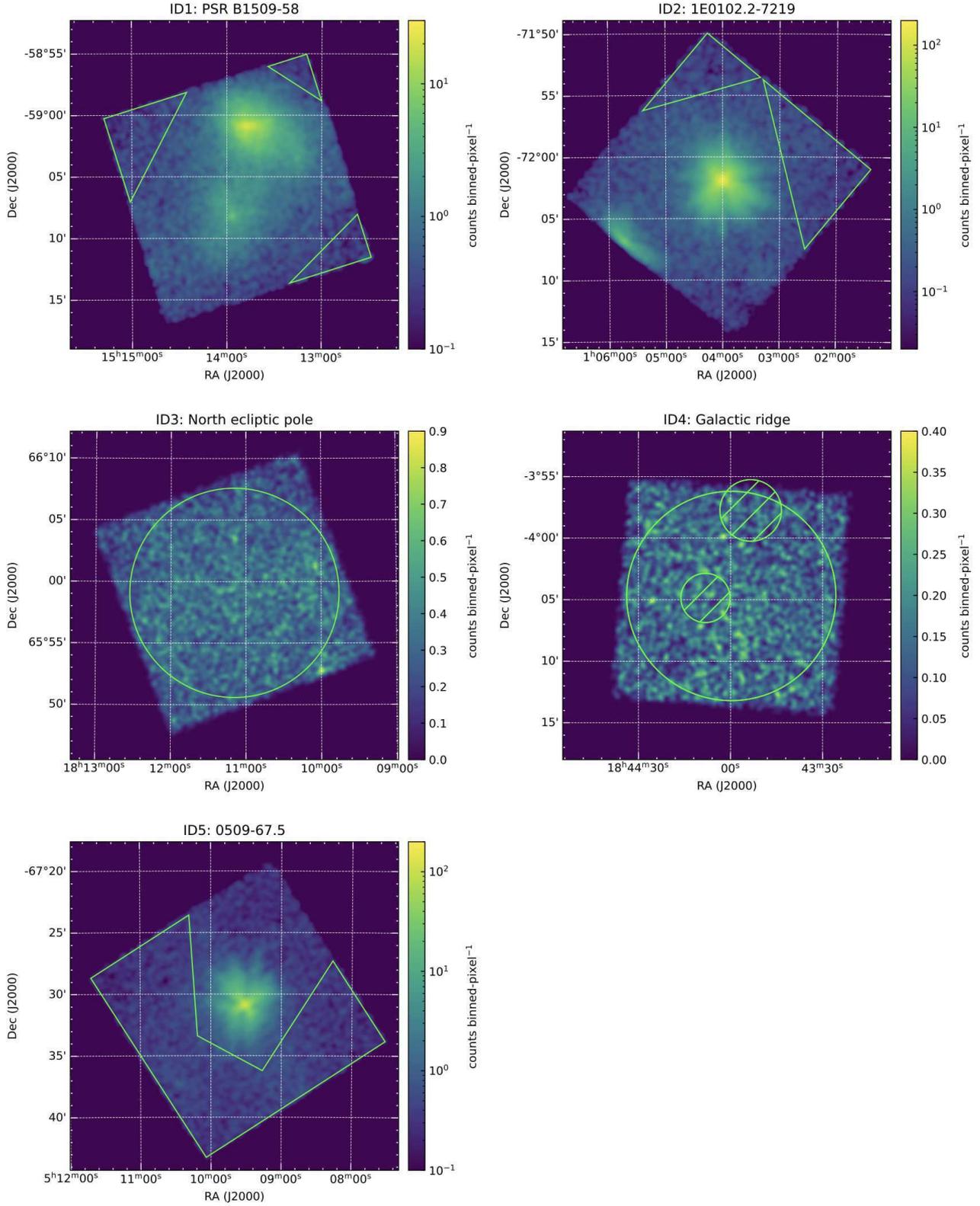} 
\end{center}
\caption{
XIS 1 0.2--1 keV images of ID1--5. 
The images are binned in 4 $\times$ 4 pixels and smoothed by a Gaussian kernel of $\sigma$ = 2.5 binned-pixels. 
The green triangle, circle, and polygon regions except for the hatched circle ones are used for 
light curve and spectral analyses. 
}
\label{fig2} 
\end{figure*}

\begin{figure*}[t]
\begin{center}
\includegraphics[width=\textwidth]{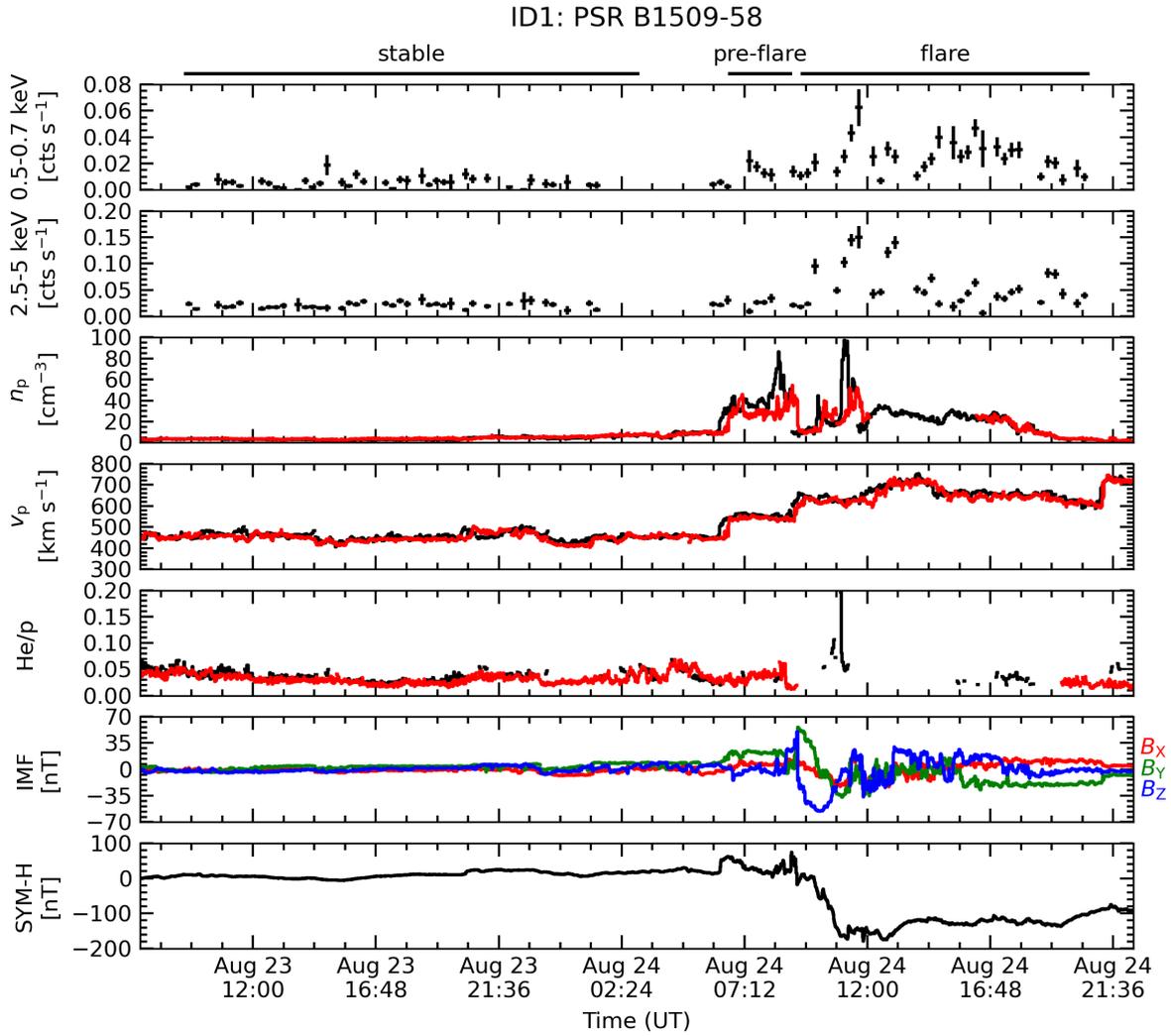} 
\end{center}
\caption{
XIS 1 0.5--0.7 keV and 2.5--5 keV light curves extracted from the TDX region of ID1, 
solar wind proton density $n_{\rm p}$, velocity $v_{\rm p}$, helium to proton ratio He/p, 
IMF $B_\mathrm{X}$, $B_\mathrm{Y}$, and $B_\mathrm{Z}$ in GSM coordinates, 
and SYM-H index as functions of times in UT. 
The vertical errors are 1$\sigma$ significance. 
The solar wind parameters were taken from the WIND and ACE satellites (black and red). 
The IMF components were taken from the ACE satellite. 
The WIND and ACE data were time-shifted to the near-Earth region. 
The SYM-H index was taken from the World Data Center for Geomagnetism, Kyoto. 
}
\label{fig3} 
\end{figure*}

\begin{figure*}[t]
\begin{center}
\includegraphics[width=\textwidth]{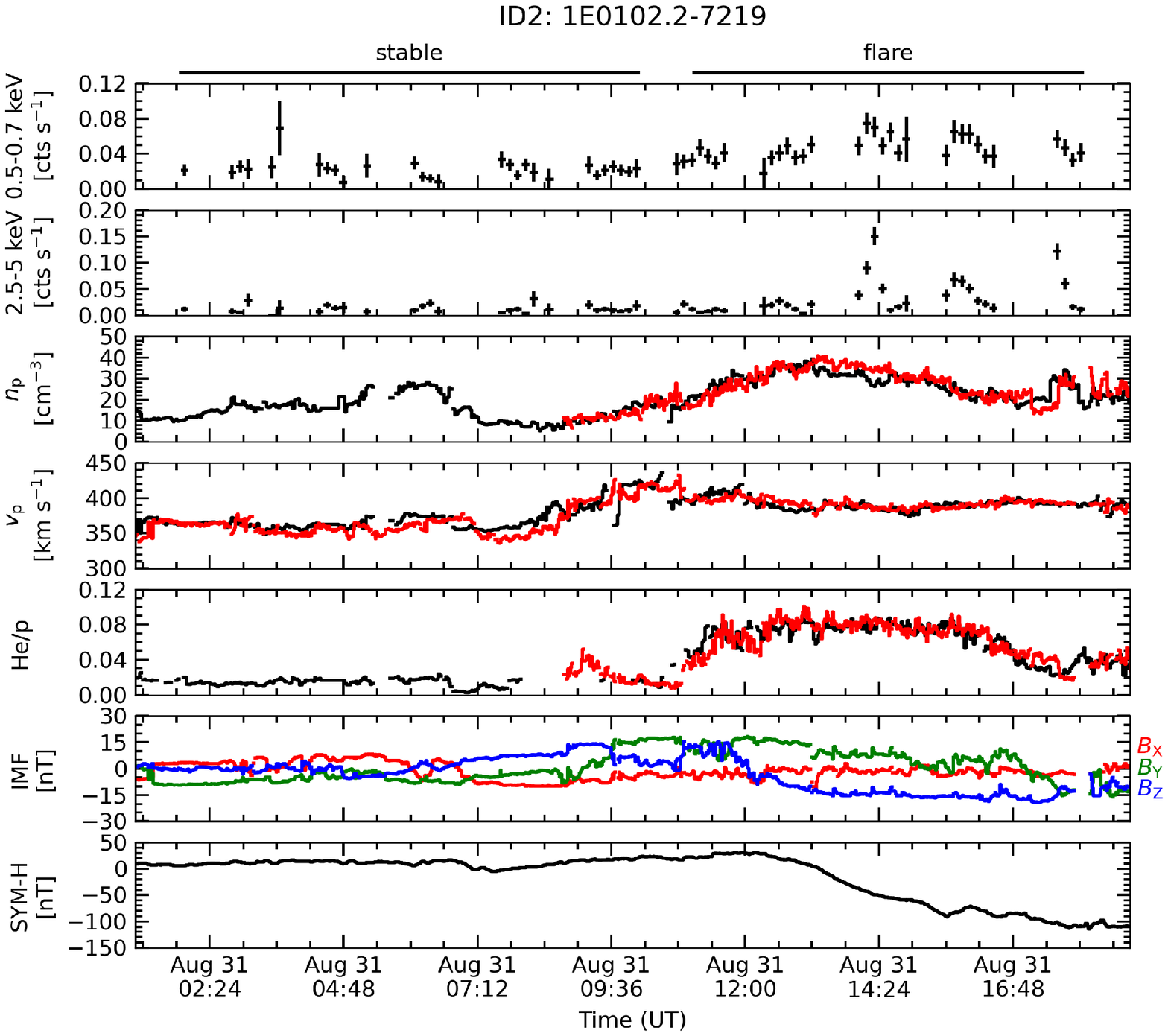} 
\end{center}
\caption{
Same as figure \ref{fig3}, but for ID2. 
}
\label{fig4} 
\end{figure*}

\begin{figure*}[t]
\begin{center}
\includegraphics[width=\textwidth]{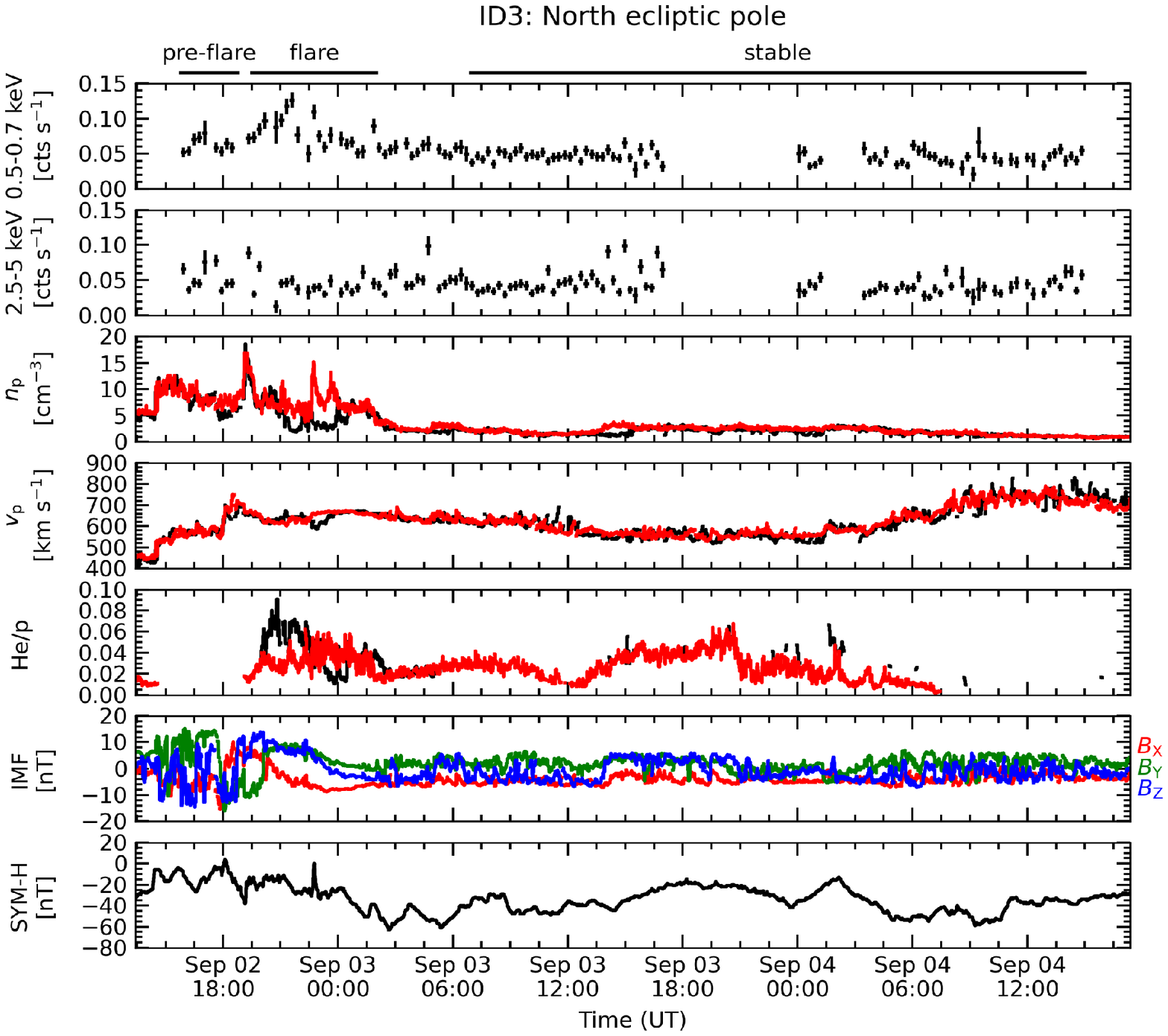} 
\end{center}
\caption{
Same as figure \ref{fig3}, but for ID3. 
}
\label{fig5} 
\end{figure*}

\begin{figure*}[t]
\begin{center}
\includegraphics[width=\textwidth]{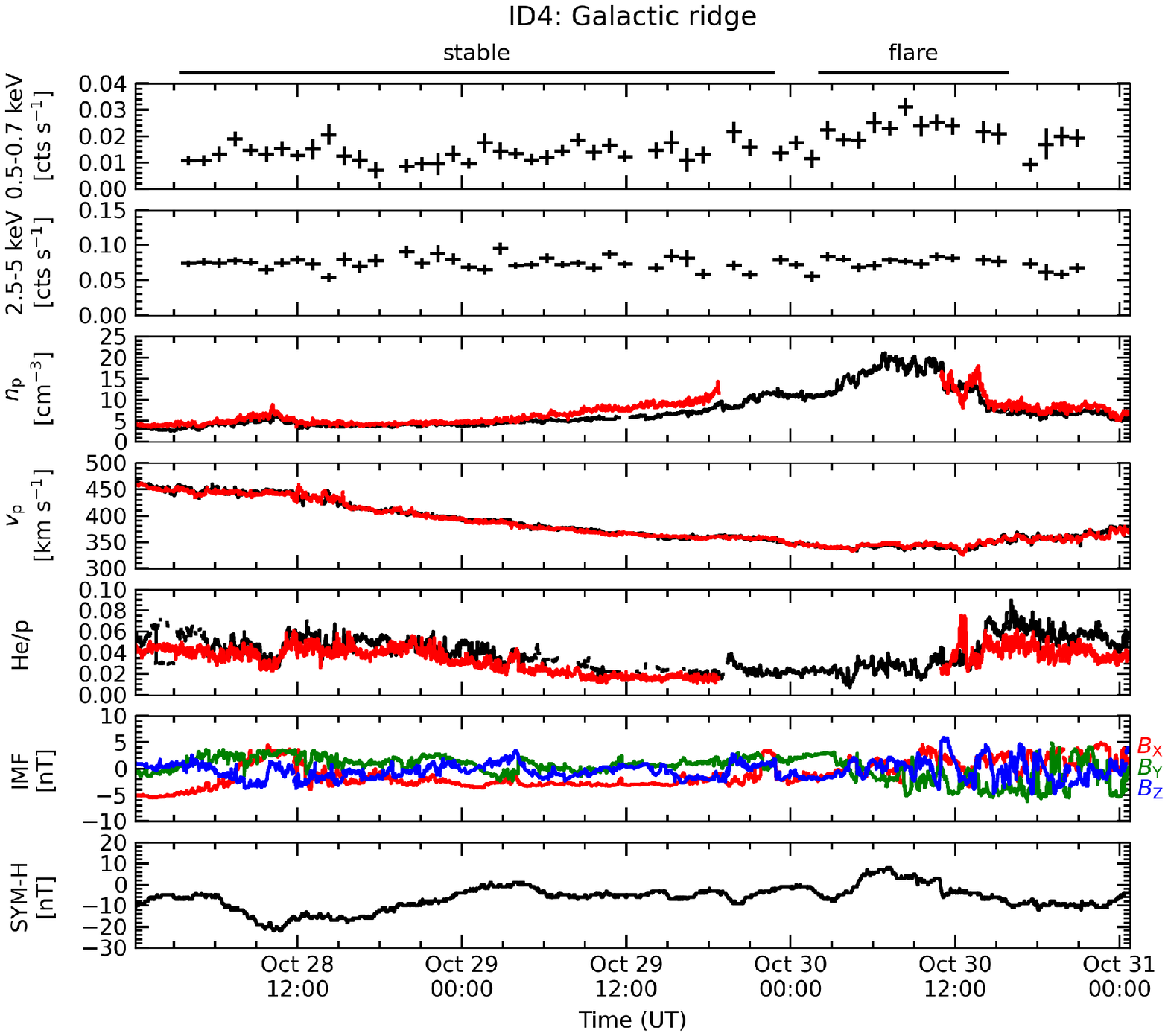} 
\end{center}
\caption{
Same as figure \ref{fig3}, but for ID4. 
}
\label{fig6} 
\end{figure*}

\begin{figure*}[t]
\begin{center}
\includegraphics[width=\textwidth]{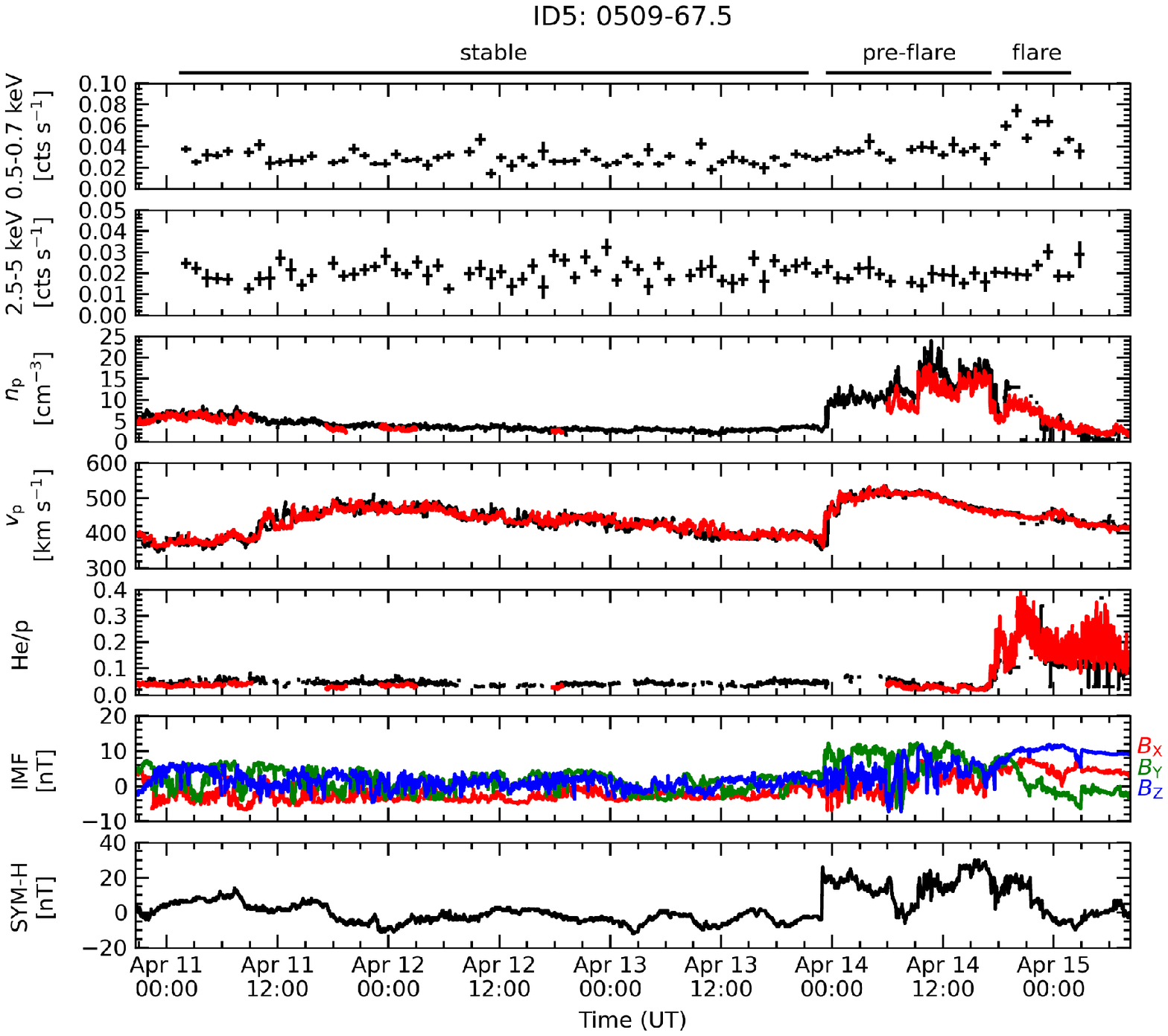} 
\end{center}
\caption{
Same as figure \ref{fig3}, but for ID5. 
}
\label{fig7} 
\end{figure*}

\begin{figure*}[t]
\begin{center}
\includegraphics[width=0.5\textwidth]{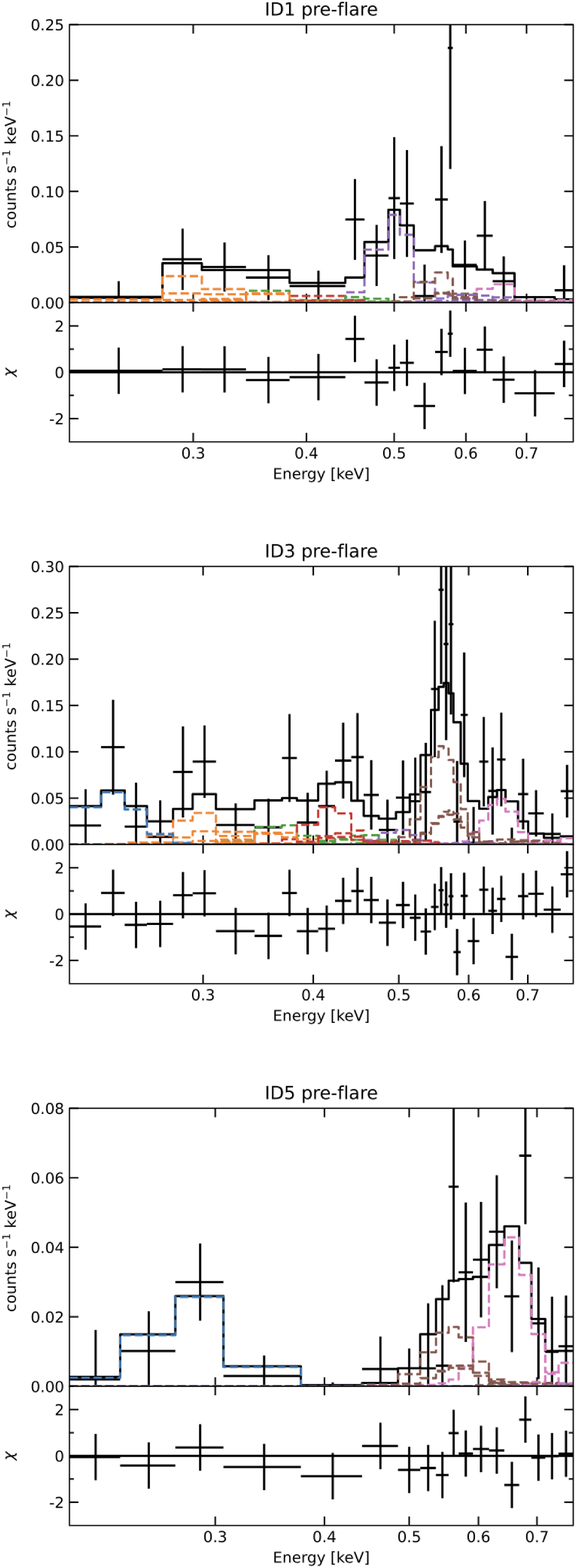} 
\end{center}
\caption{
XIS 1 spectra during the {\it pre-flare} period of ID1, 3, and 5. 
The {\it stable} spectrum during each observation is subtracted as a background. 
The Bodewits model and a narrow Gaussian reproducing the lowest energy line are used. 
Their parameters are listed in table \ref{tab3}. 
}
\label{fig8} 
\end{figure*}

\begin{figure*}[t]
\begin{center}
\includegraphics[width=0.5\textwidth]{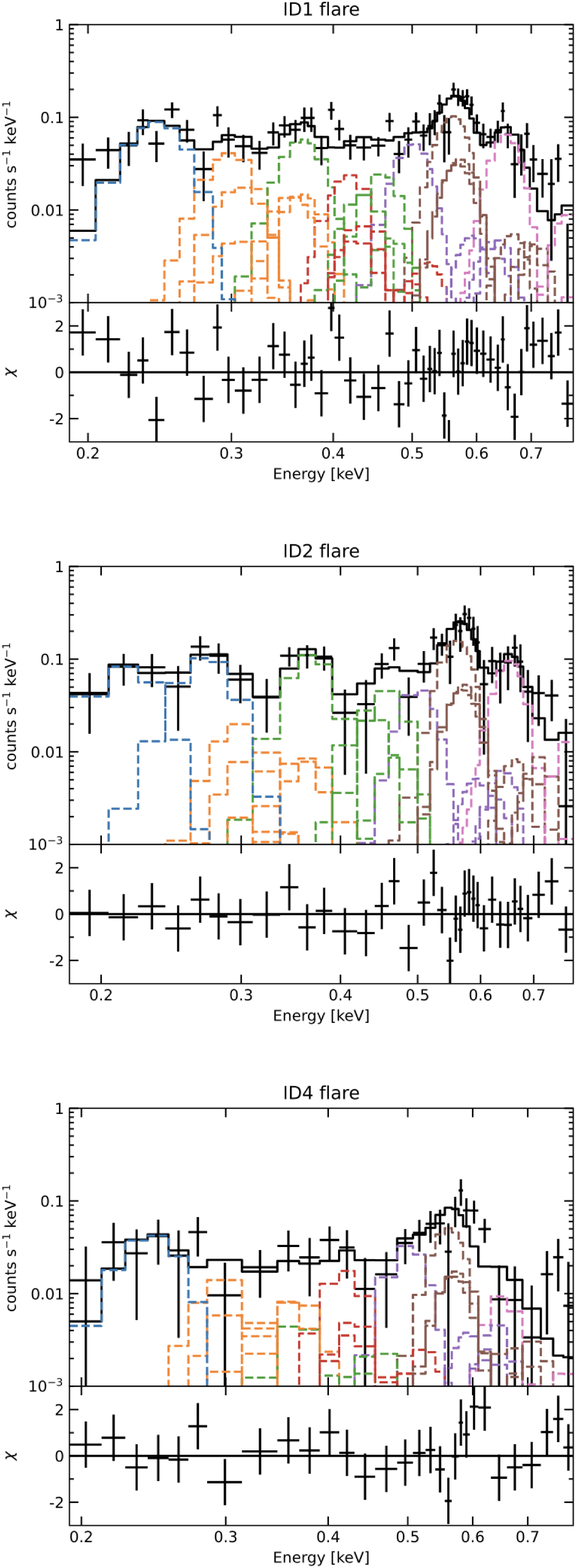} 
\end{center}
\caption{
XIS 1 spectra during the {\it flare} period of ID1, 2, and 4. 
The {\it stable} spectrum during each observation is subtracted as a background. 
The Bodewits model and one or two narrow Gaussians reproducing the lowest energy lines are used. 
Their parameters are listed in table \ref{tab4}. 
}
\label{fig9} 
\end{figure*}

\begin{figure*}[t]
\begin{center}
\includegraphics[width=\textwidth]{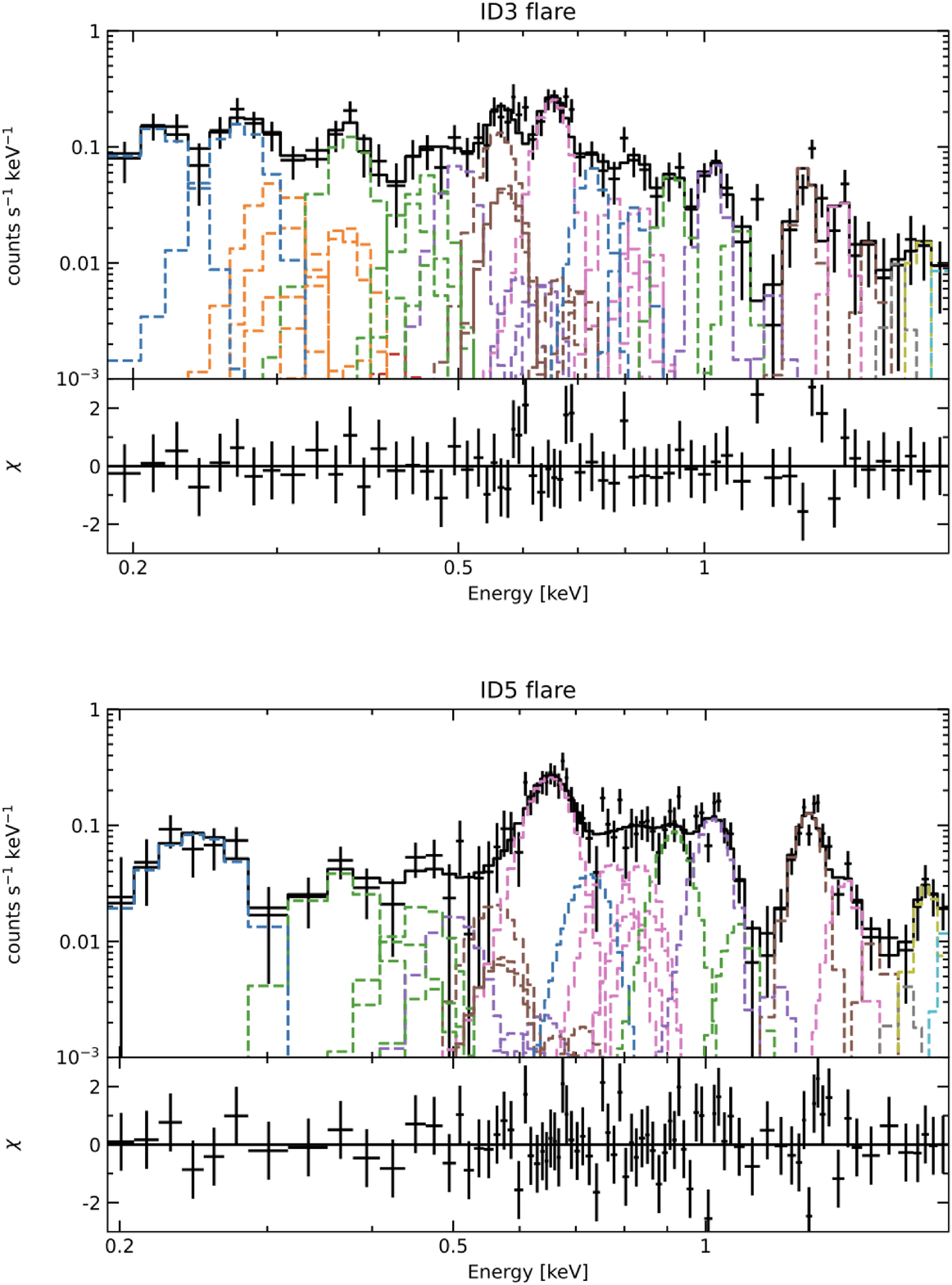} 
\end{center}
\caption{
XIS 1 spectra during the {\it flare} period of ID3 and 5. 
The {\it stable} spectrum during each observation is subtracted as a background. 
The Bodewits model, one or two narrow Gaussians reproducing the lowest energy lines, and 
14 narrow Gaussians reproducing emission lines at higher energies are used. 
Their parameters are listed in table \ref{tab4}. 
}
\label{fig10} 
\end{figure*}

\begin{figure*}[t]
\begin{center}
\includegraphics[width=\textwidth]{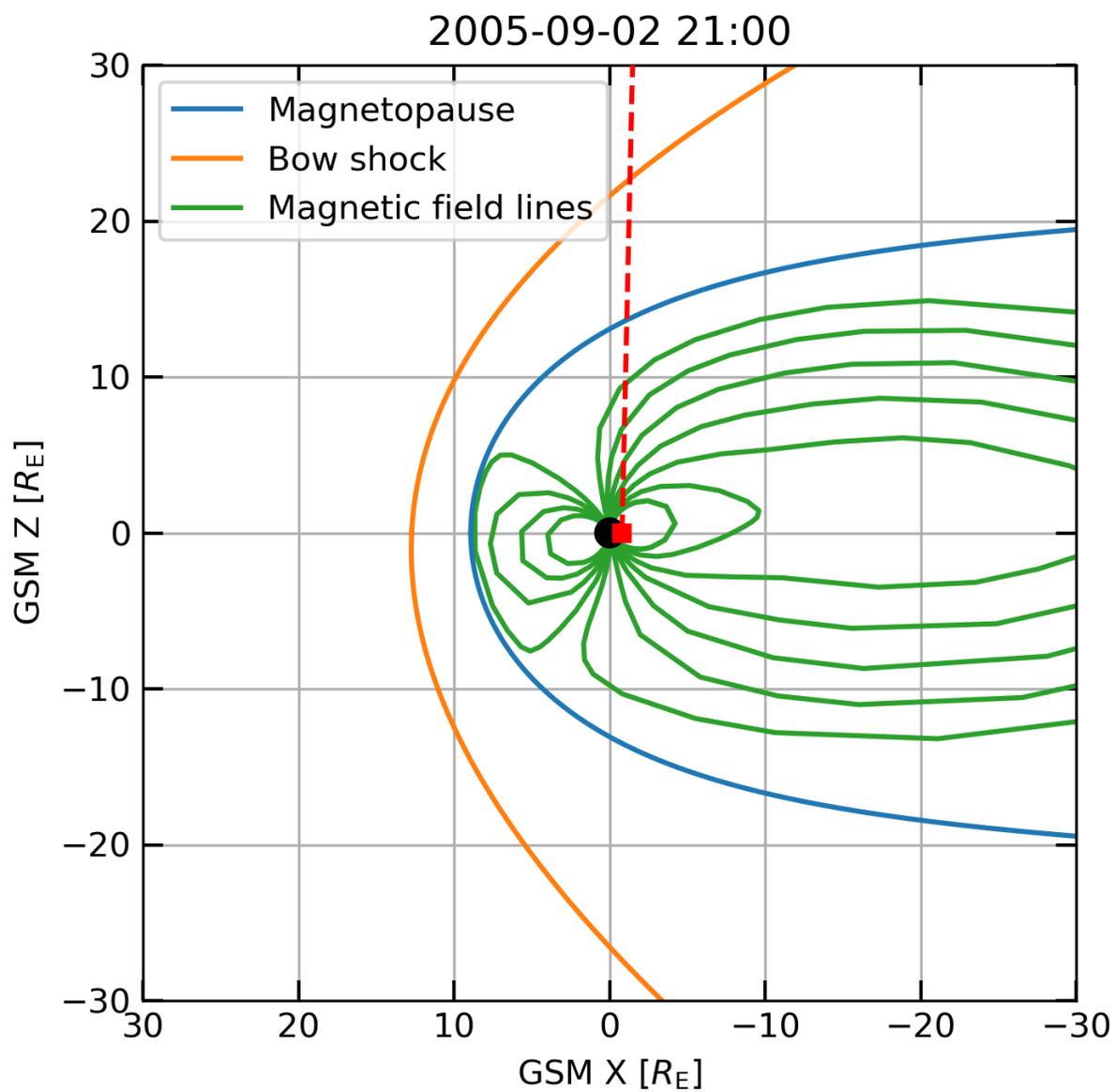} 
\end{center}
\caption{
Example of modeled magnetospheric configuration in GSM XZ plane. 
The blue, orange, and green lines indicate magnetopause, bow shock, and magnetic field lines. 
The red square and dotted line represent Suzaku position and line-of-sight direction.
} 
\label{fig11} 
\end{figure*}

\begin{figure*}[t]
\begin{center}
\includegraphics[width=\textwidth]{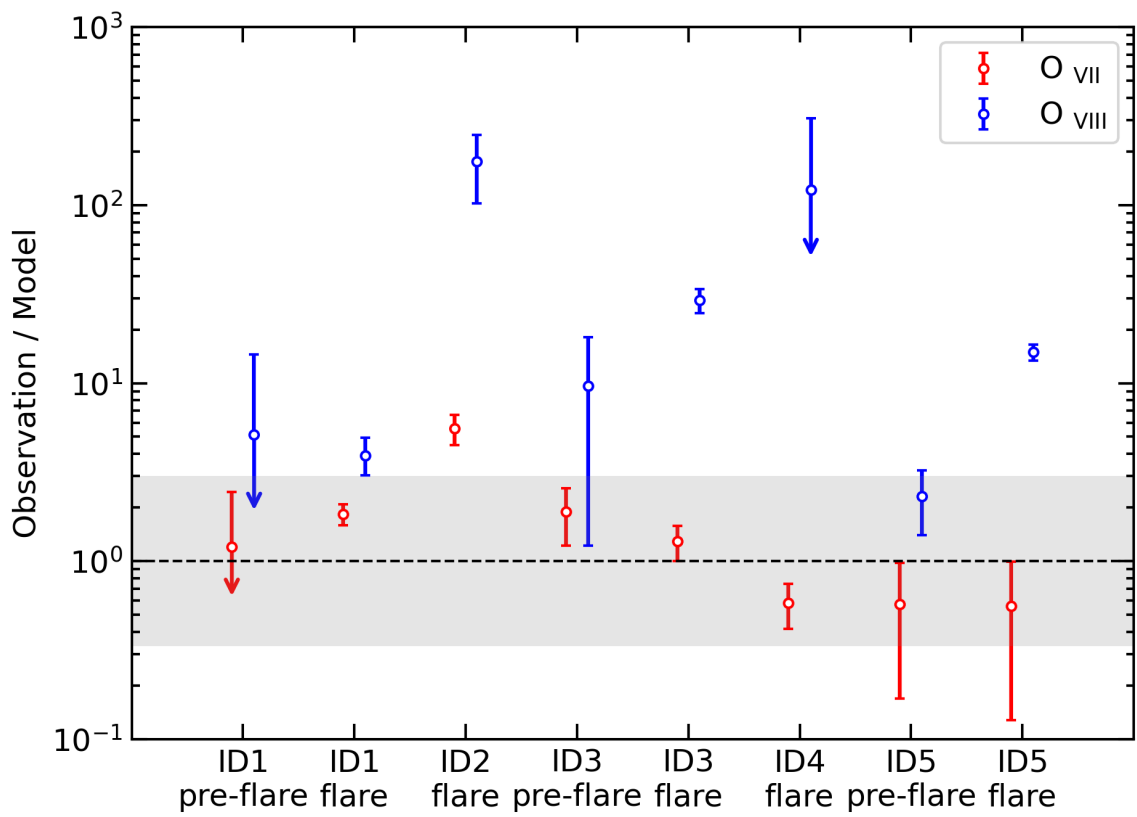} 
\end{center}
\caption{
Ratio of observation to model for O\,\emissiontype{VII} and O\,\emissiontype{VIII} line fluxes. 
The black shaded area indicates a ratio within a factor of three.
}
\label{fig12} 
\end{figure*}

\begin{figure*}[t]
\begin{center}
\includegraphics[width=\textwidth]{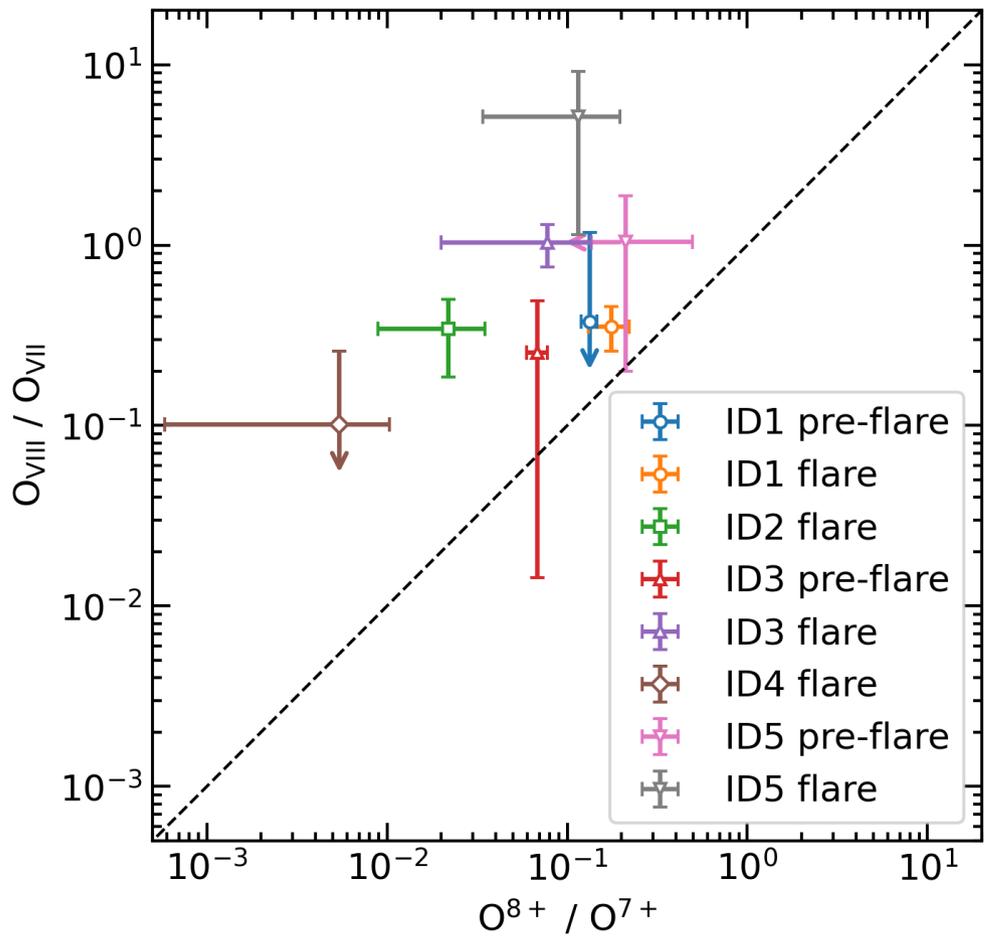} 
\end{center}
\caption{
Comparison between O$^{8+}$/O$^{7+}$ ion ratio measured by ACE/SWICS and 
O\,\emissiontype{VIII}/O\,\emissiontype{VII} flux ratio deduced from geocoronal SWCX spectra. 
}
\label{fig13} 
\end{figure*}

\begin{figure*}[t]
\begin{center}
\includegraphics[width=\textwidth]{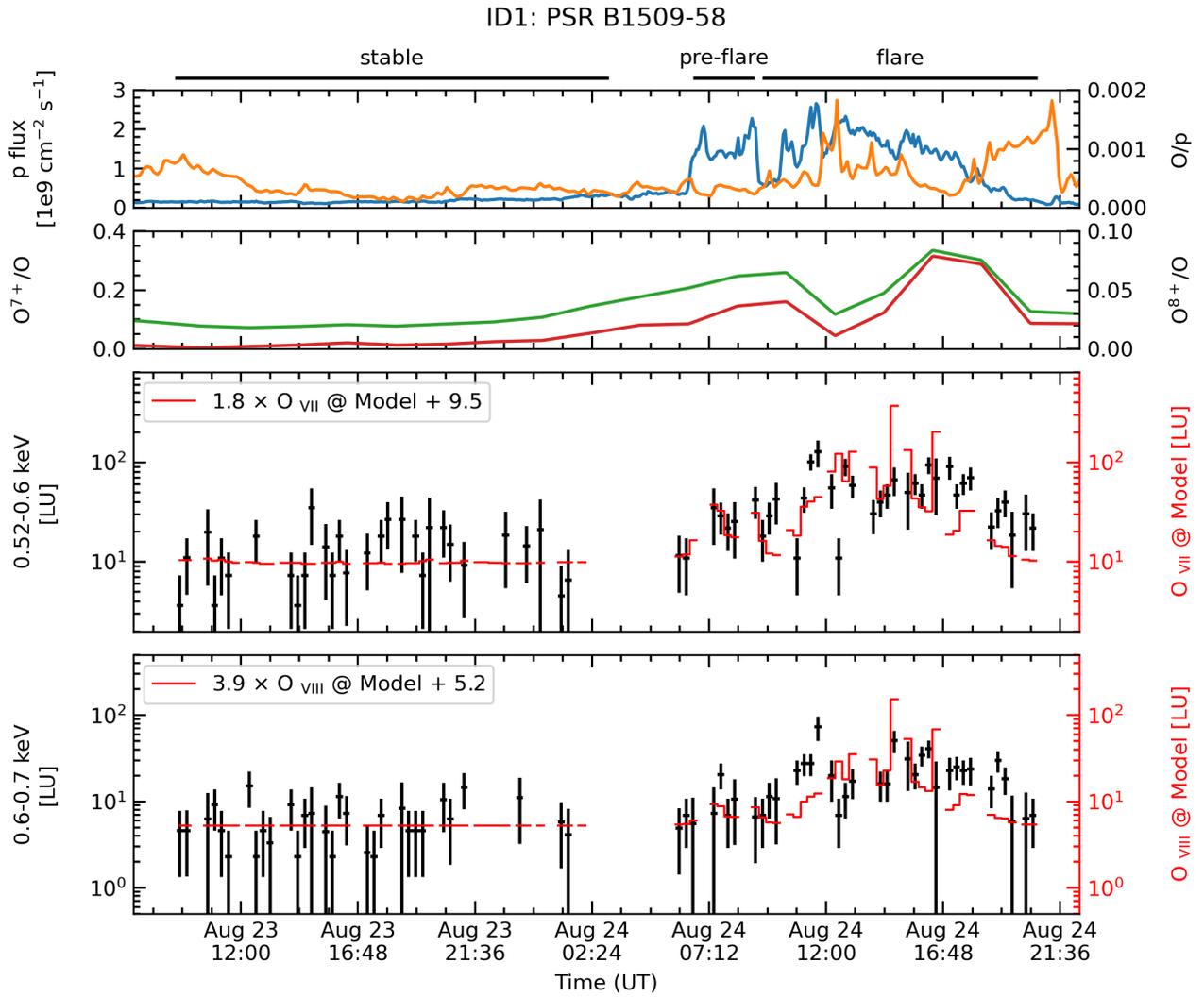} 
\end{center}
\caption{
Solar wind proton flux (blue), oxygen to proton ratio (orange), 
oxygen charge state fractions for O$^{7+}$ (green) and O$^{8+}$ ions (red), 
XIS 1 0.52--0.6 keV and 0.6--0.7 keV light curves extracted from the TDX region of ID1 (black), and 
model light curves of ID1 for O\,\emissiontype{VII} and O\,\emissiontype{VIII} emission lines (red). 
The numbers in boxes indicate scaling factors and background levels in units of LU.
}
\label{fig14} 
\end{figure*}

\begin{figure*}[t]
\begin{center}
\includegraphics[width=\textwidth]{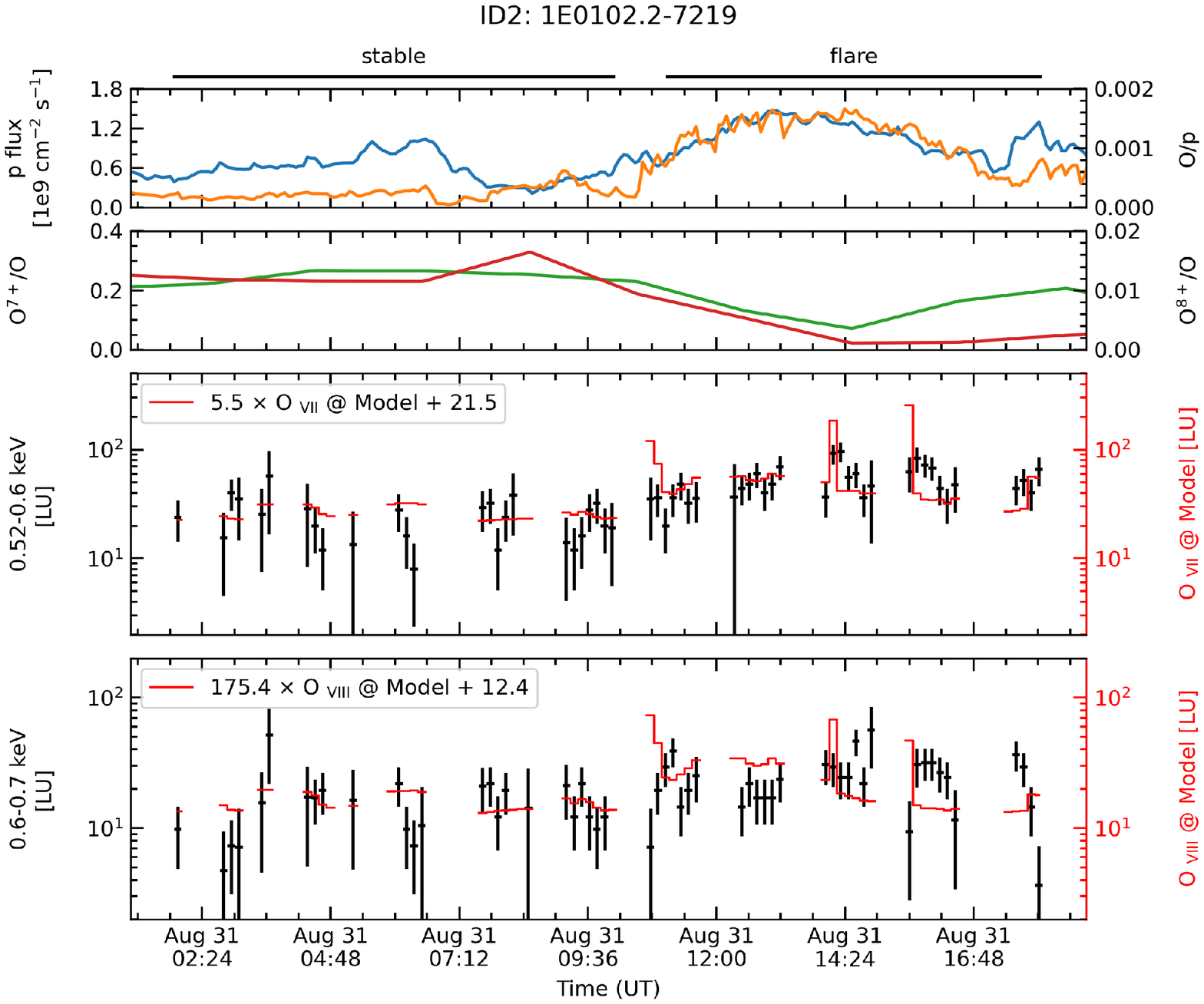} 
\end{center}
\caption{
Same as figure \ref{fig14}, but for ID2.
}
\label{fig15} 
\end{figure*}

\begin{figure*}[t]
\begin{center}
\includegraphics[width=\textwidth]{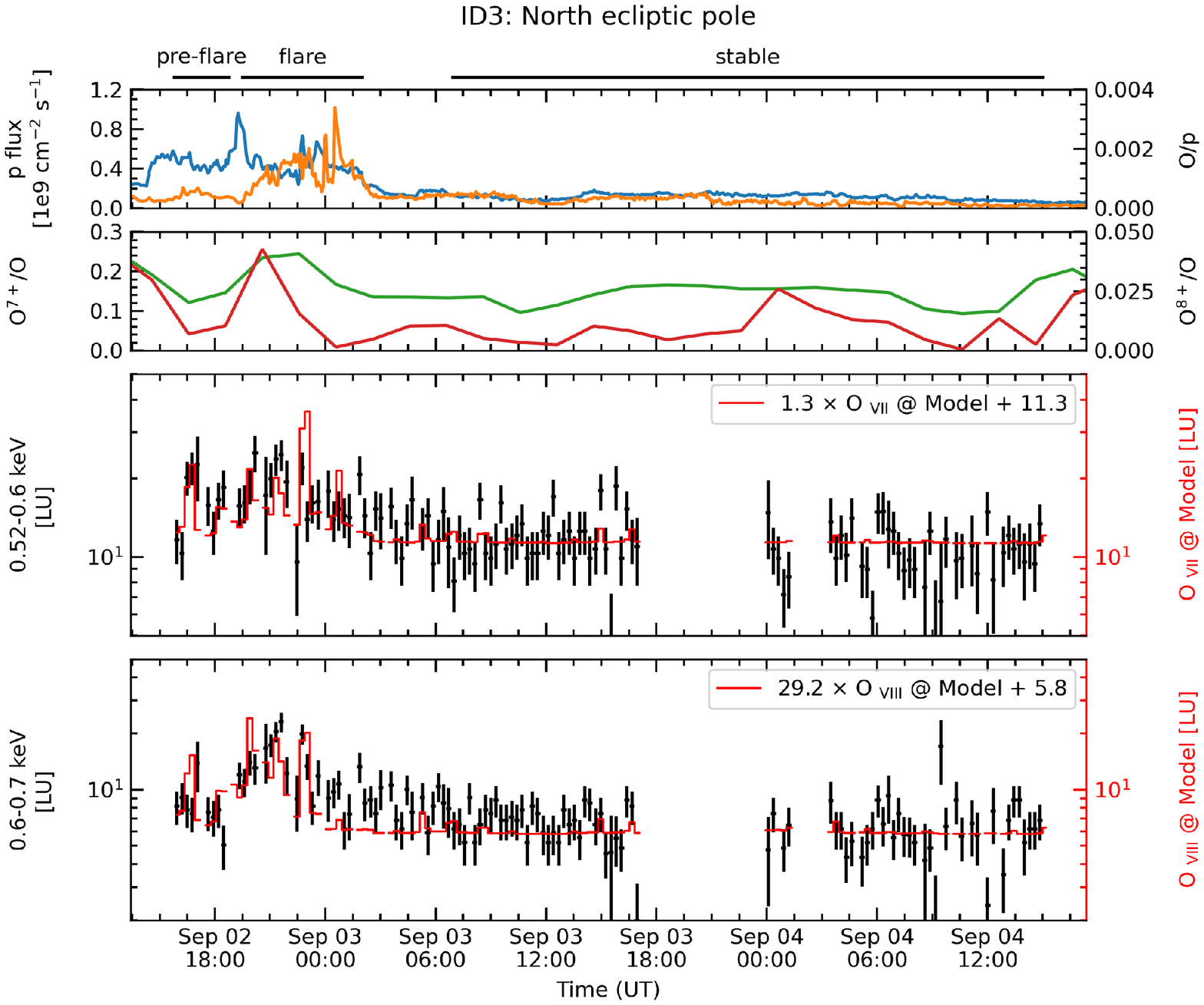} 
\end{center}
\caption{
Same as figure \ref{fig14}, but for ID3. 
}
\label{fig16} 
\end{figure*}

\begin{figure*}[t]
\begin{center}
\includegraphics[width=\textwidth]{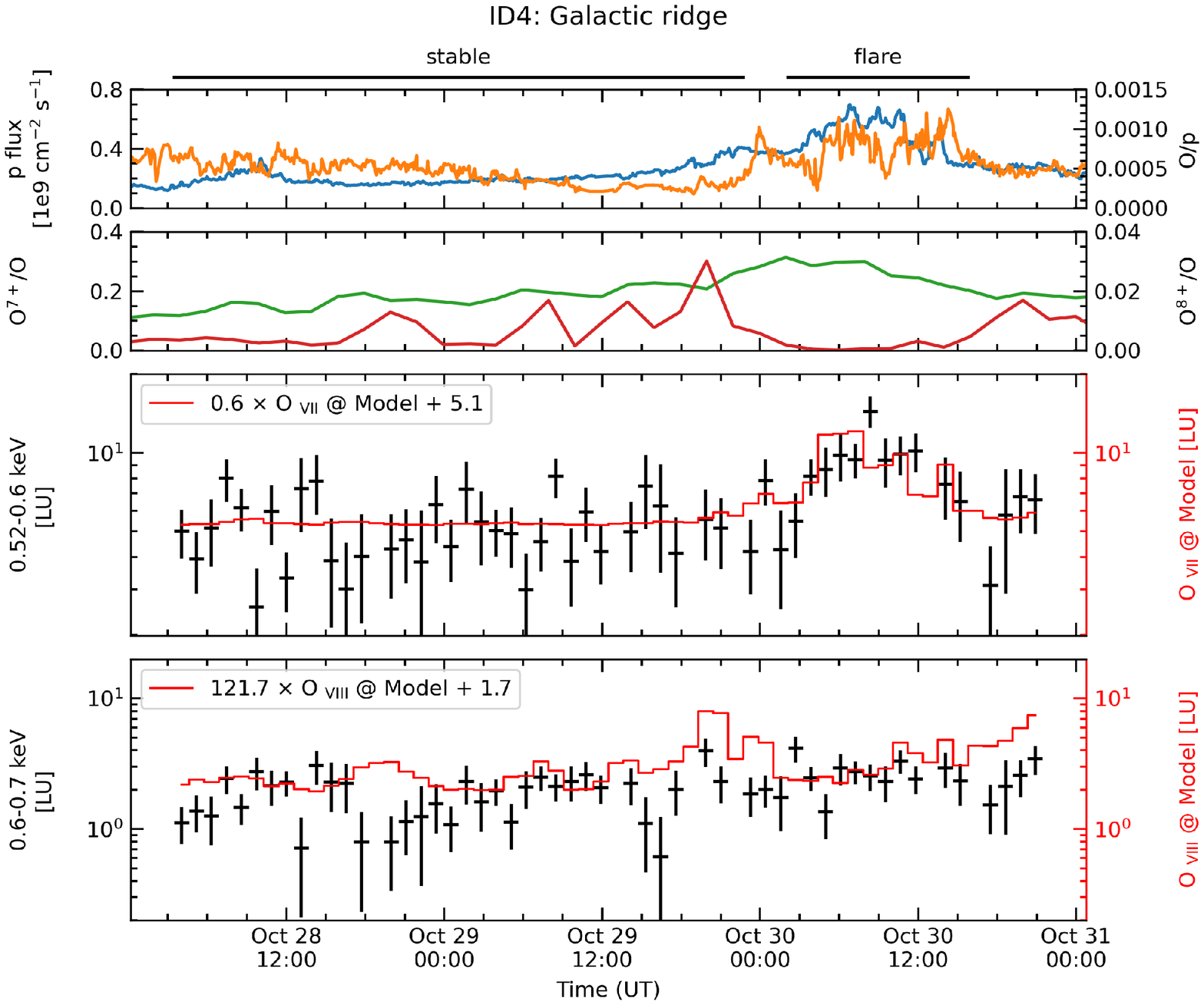} 
\end{center}
\caption{
Same as figure \ref{fig14}, but for ID4. 
}
\label{fig17} 
\end{figure*}

\begin{figure*}[t]
\begin{center}
\includegraphics[width=\textwidth]{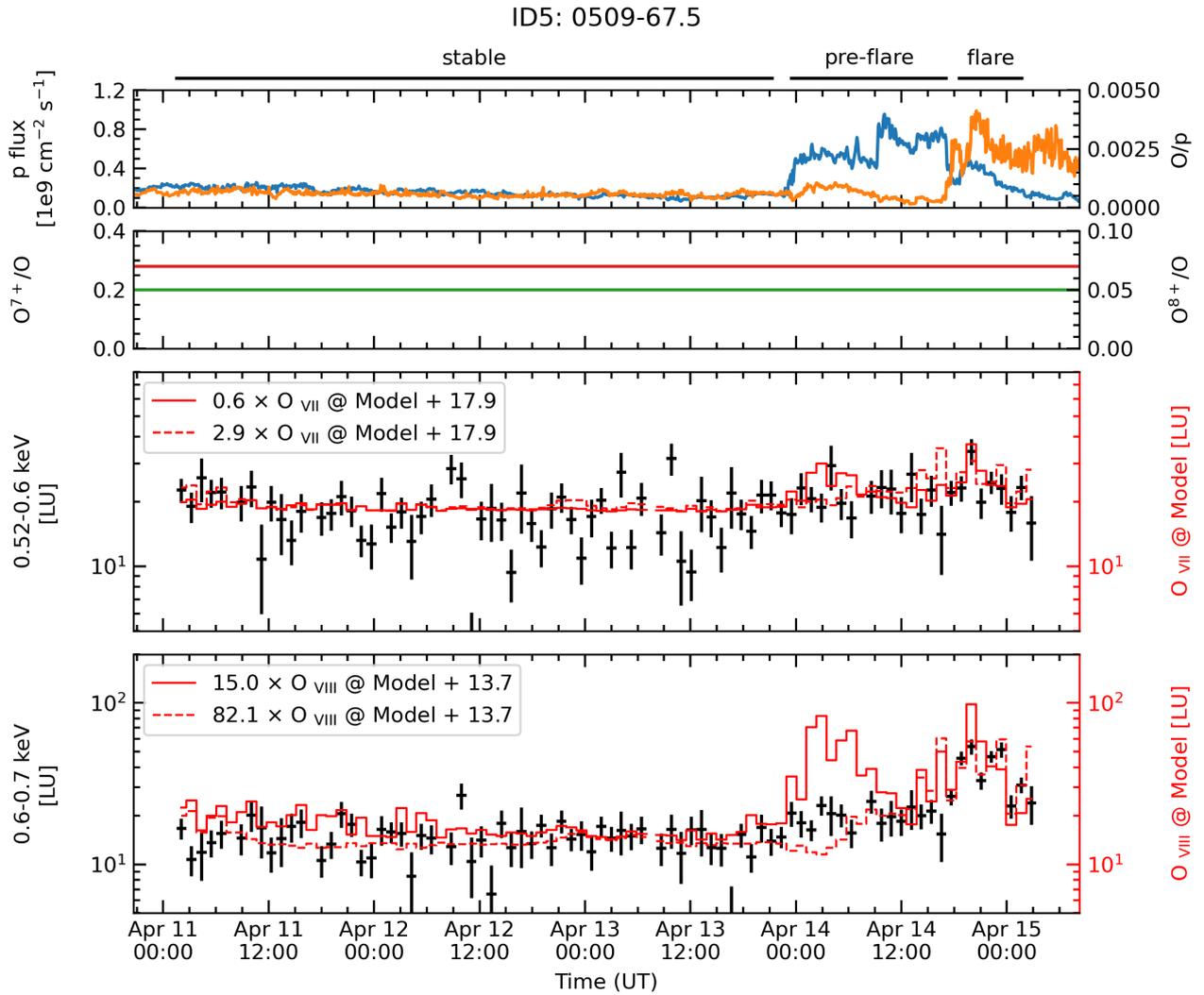} 
\end{center}
\caption{
Same as figure \ref{fig14}, but for ID5. 
Note that the solid and dotted lines represent model light curves using constant oxygen charge state fractions 
and time-variable oxygen ion fluxes deduced from an empirical equation (see text). 
}
\label{fig18} 
\end{figure*}

\begin{figure*}[t]
\begin{center}
\includegraphics[width=\textwidth]{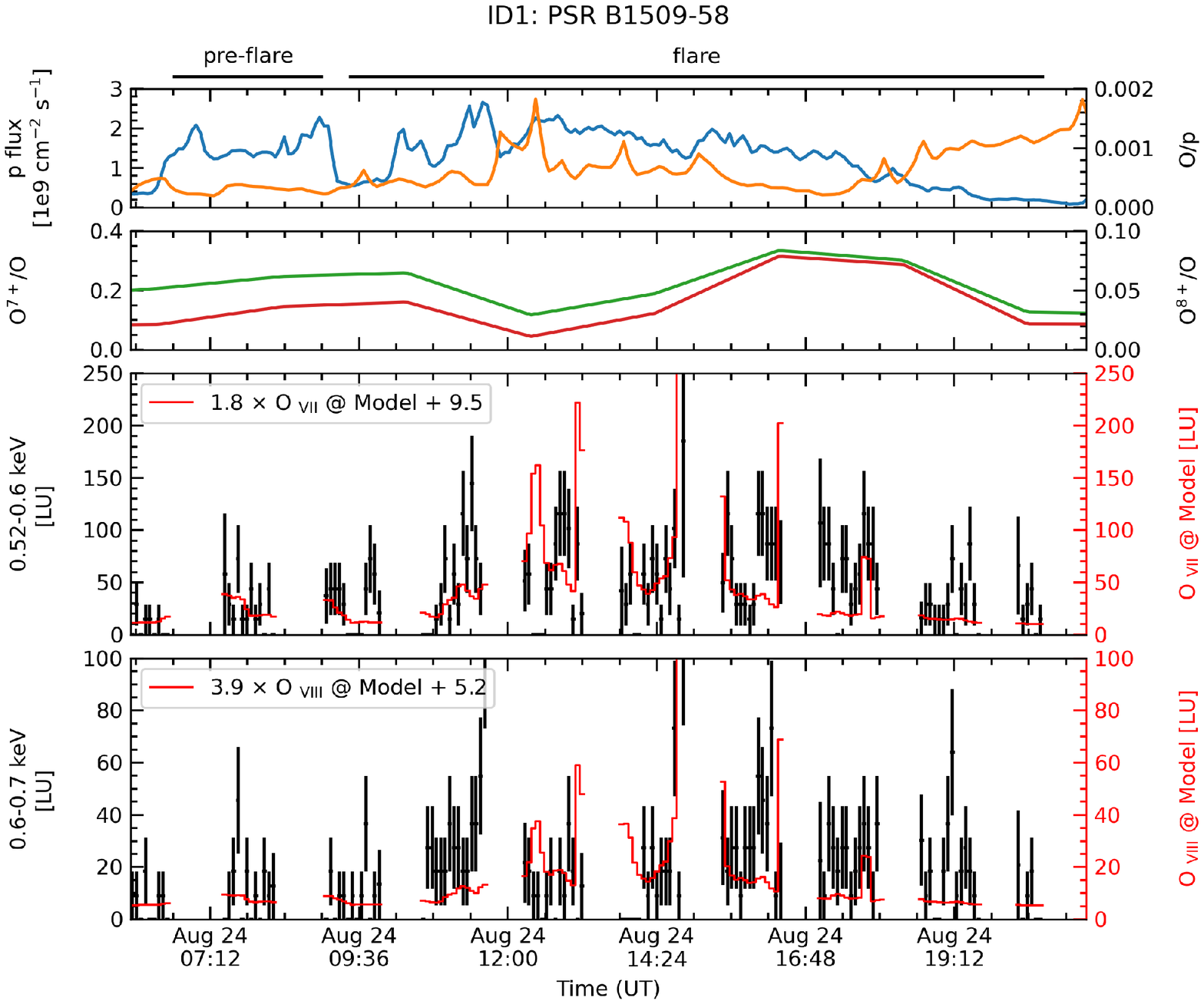} 
\end{center}
\caption{
Enlarged view during the {\it pre-flare} and {\it flare} periods of figure \ref{fig14}. 
The time bin is much shorter. 
}
\label{fig19} 
\end{figure*}

\begin{figure*}[t]
\begin{center}
\includegraphics[width=\textwidth]{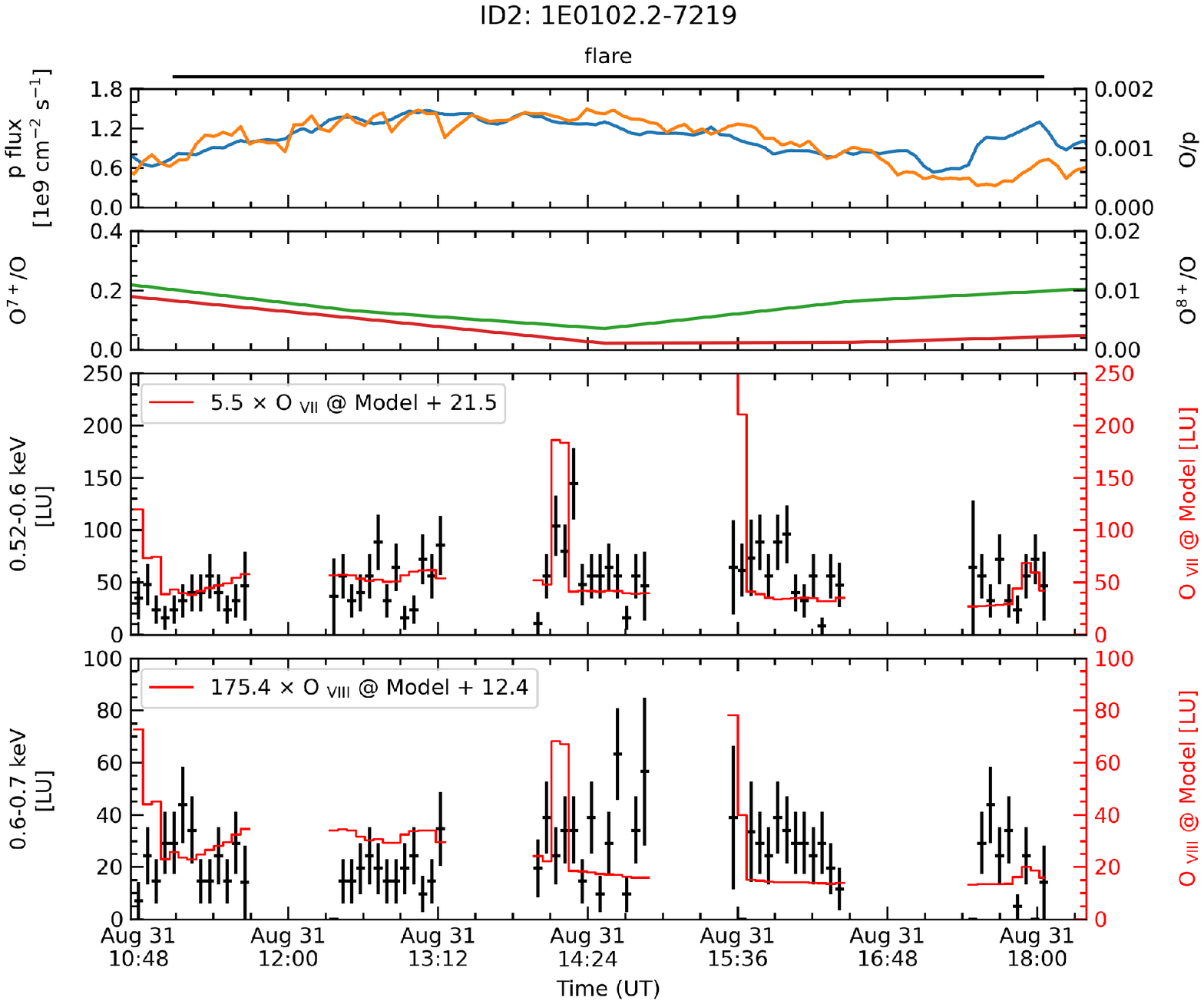} 
\end{center}
\caption{
Enlarged view during the {\it flare} period of figure \ref{fig15}. 
The time bin is much shorter. 
}
\label{fig20} 
\end{figure*}

\begin{figure*}[t]
\begin{center}
\includegraphics[width=\textwidth]{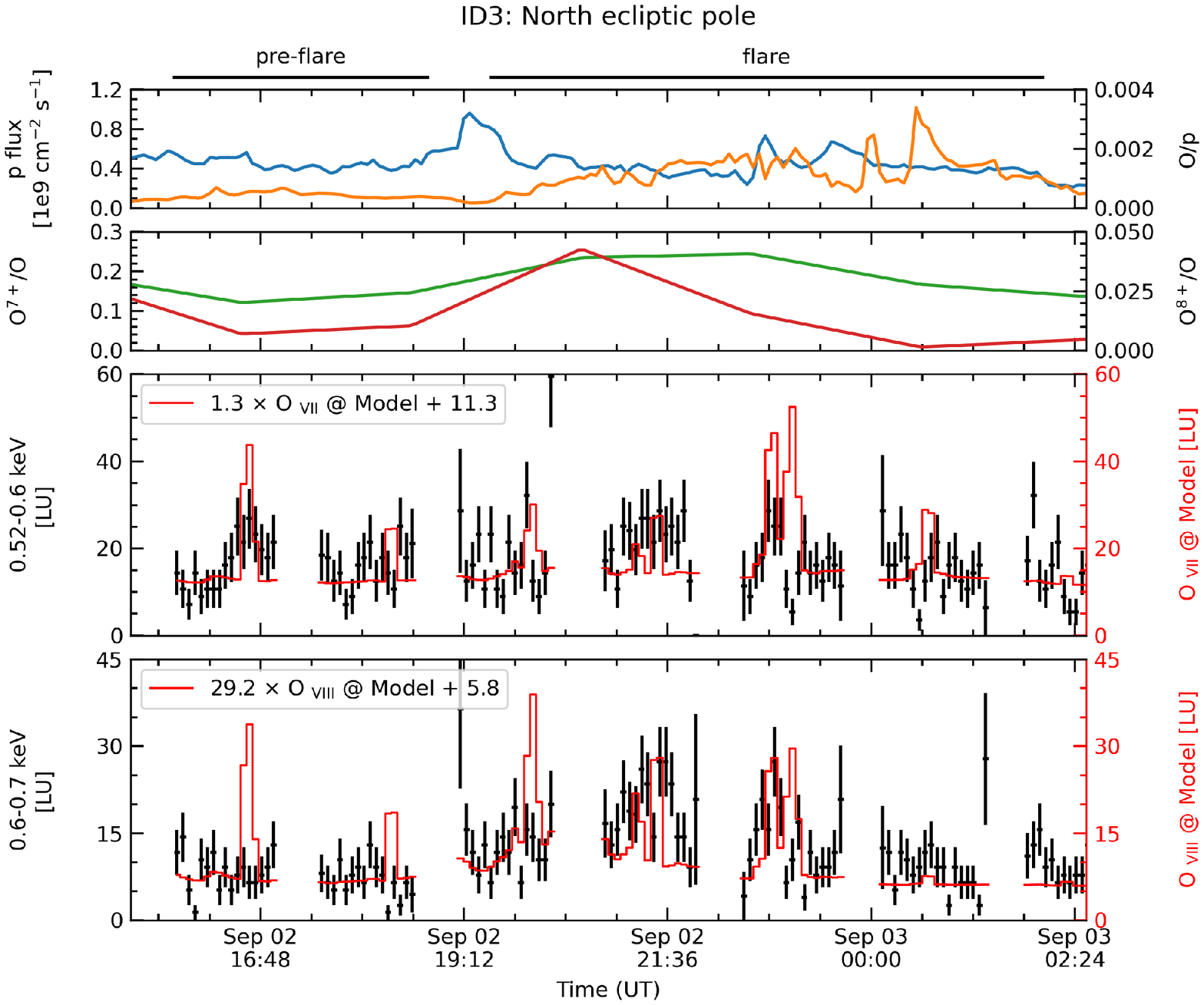} 
\end{center}
\caption{
Enlarged view during the {\it pre-flare} and {\it flare} periods of figure \ref{fig16}. 
The time bin is much shorter. 
}
\label{fig21} 
\end{figure*}

\begin{figure*}[t]
\begin{center}
\includegraphics[width=\textwidth]{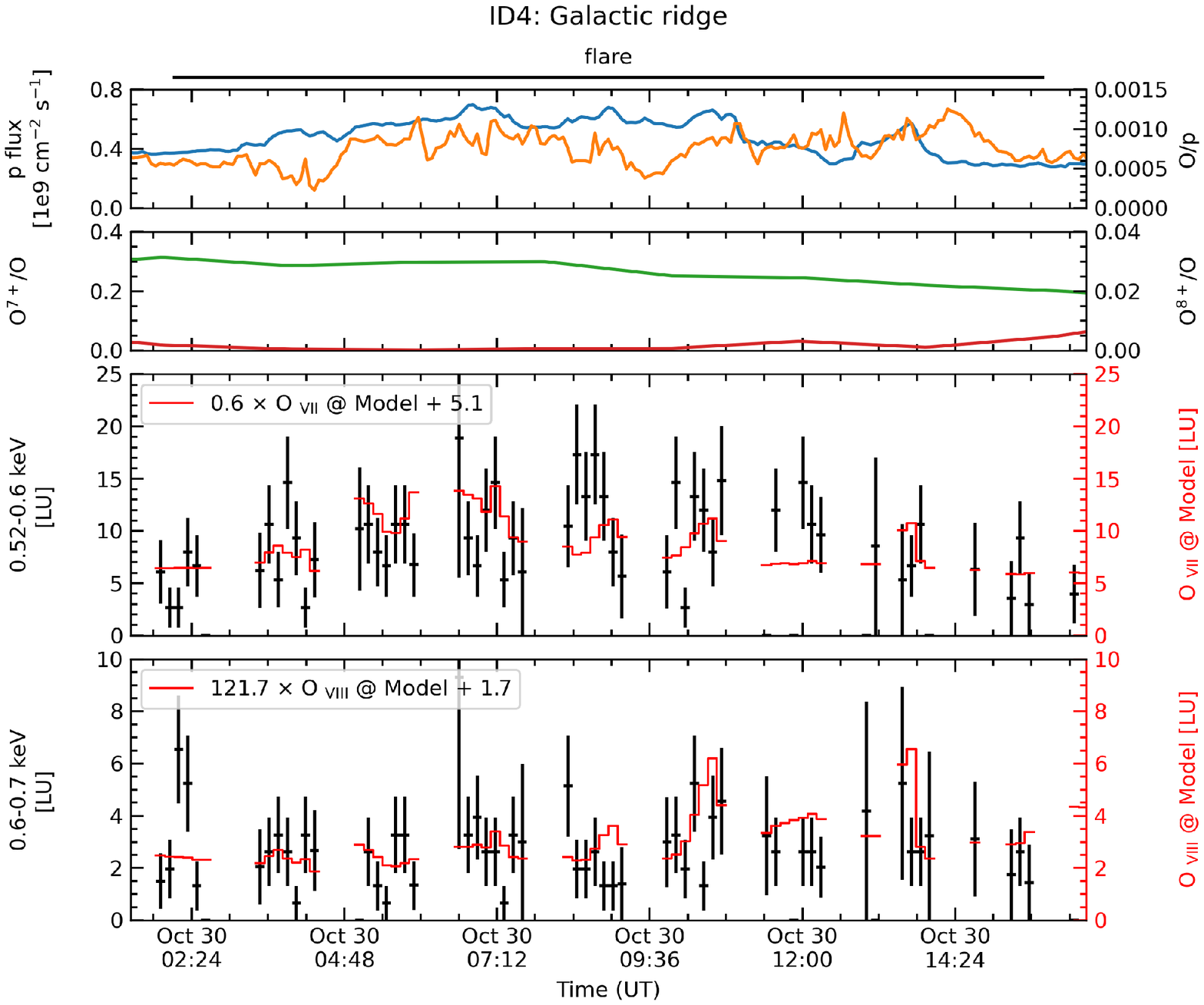} 
\end{center}
\caption{
Enlarged view during the {\it flare} period of figure \ref{fig17}. 
The time bin is much shorter. 
}
\label{fig22} 
\end{figure*}

\begin{figure*}[t]
\begin{center}
\includegraphics[width=\textwidth]{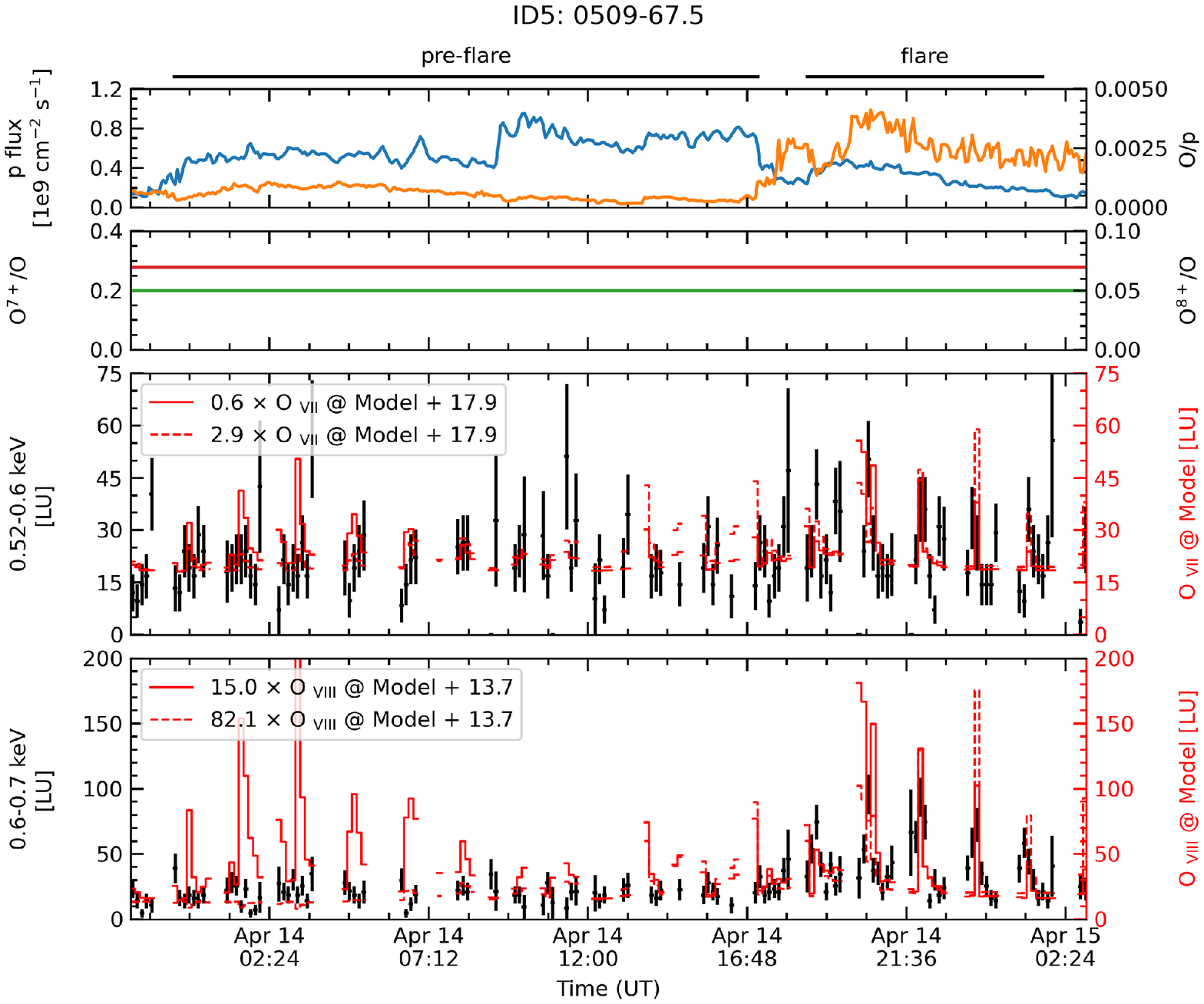} 
\end{center}
\caption{
Enlarged view during the {\it pre-flare} and {\it flare} periods of figure \ref{fig18}. 
The time bin is much shorter. 
}
\label{fig23} 
\end{figure*}


\end{document}